# A spontaneously patterning reaction–diffusion network, containing an integrated activator–inhibitor and substrate depletion mechanism, specifies trichoblast cell fate in *Arabidopsis* roots

## Author list

Hayley Mills[1], George Janes[2], Anthony Bishopp[2], Natasha Savage[1]*

[1]Institute of Systems, Molecular and Integrative Biology. University of Liverpool. UK.

[2]School of Biosciences. University of Nottingham. UK.

*nsavage@liverpool.ac.uk

## Abstract

*Arabidopsis* root hair patterning is controlled by a complex transcription factor network containing positive and negative feedback loops, epidermal cell-cell signalling, and positional signalling from underlying tissue. Recently, several long accepted regulatory interactions within the network have been revised, and while there are extensive data regarding individual components, the complexity of the network has made it difficult to understand how these components combine to ensure correct and robust epidermal patterning. Here, mathematical modelling was used to integrate the wealth of experimental data into a single transcription factor network model.

Current understanding of the epidermal patterning network was found to be insufficient to reproduce experimental data, and thus an additional negative feedback loop was hypothesized which enabled the model to reproduce both wildtype and mutant data. The negative feedback was supported by sequence analysis of candidate regulators. Modelling investigations uncovered interactions, mechanisms, and constraints essential for patterning, and revealed how a recently redefined reaction functions to produce mutant data while contributing to network robustness in wildtype. When analysed together, these results provide a holistic understanding of epidermal cell fate determination in *Arabidopsis*, shown here to be governed by a spontaneously patterning reaction-diffusion network containing combined activator-inhibitor and substrate depletion mechanisms.

## Introduction

Root hairs increase the absorptive surface area of plant roots and are essential for water and nutrient uptake. On immergence from the root meristem, epidermal cells differentiate into trichoblast cells, which form a hair, or atrichoblast cells, which do not. In the model organism *Arabidopsis*, root epidermal cells are arranged in longitudinal cell files and trichoblast patterning is well described, with trichoblast cells forming in epidermal cell files that overlie cortical clefts, the junctions between adjacent cortical cell files[1,2], Fig. 1a. Epidermal cells overlying cortical clefts are said to be in H-positions, while cells lacking cortical cleft contact occupy N-positions. Trichoblast cell files forming over cortical clefts has led to the hypothesis that a signal moves through the cortical cleft and is preferentially received by H-positioned cells, biasing them towards trichoblast cell fate[3]. Despite the simplicity of the trichoblast pattern, epidermal cell fate specification is not governed by a linear signalling cascade. Instead, the underlying transcription factor network contains both positive and negative feedback loops as well as epidermal cell-cell signalling, Fig. 1b.

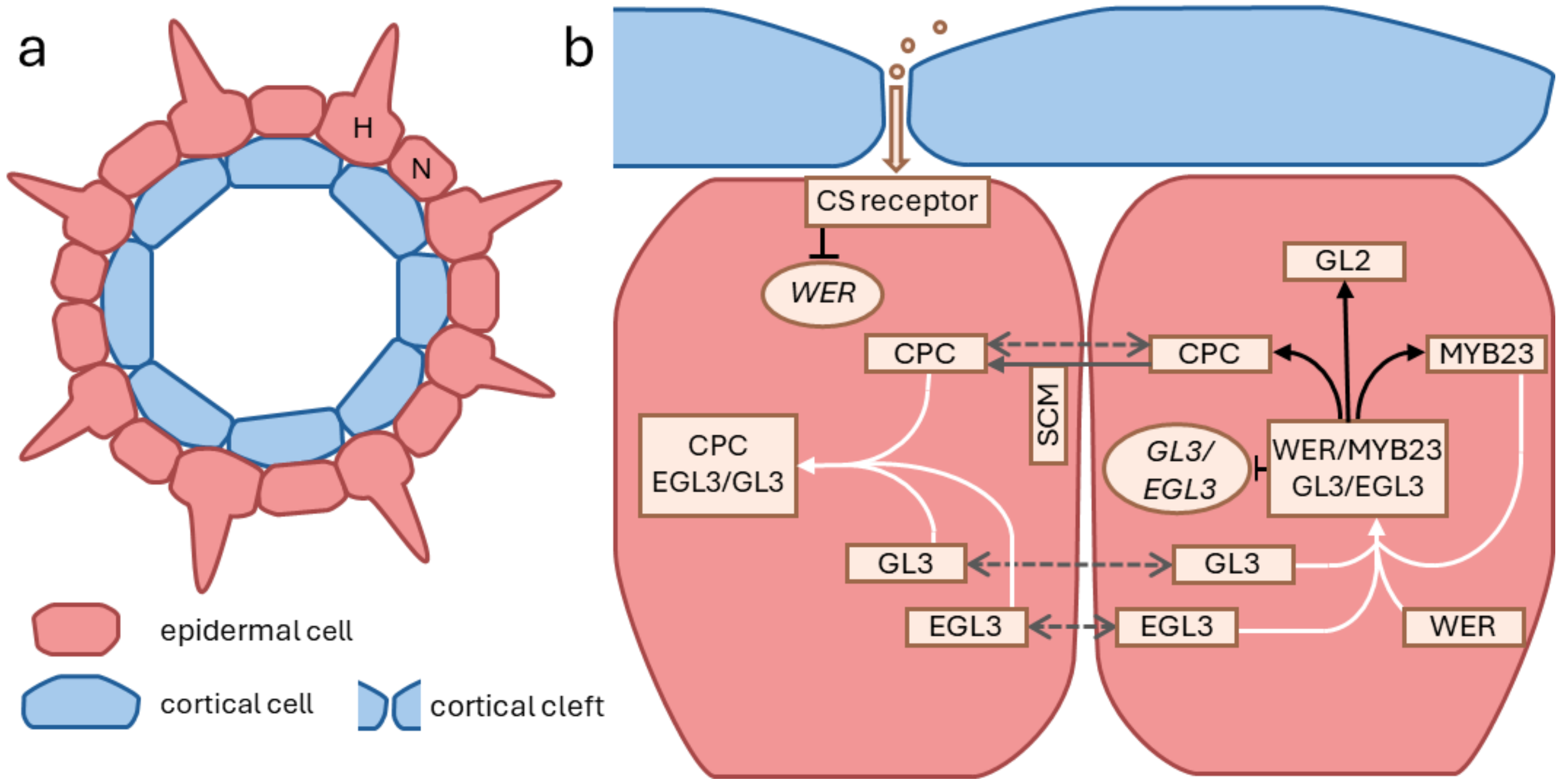

**Figure 1: Epidermal and cortical layers in the *Arabidopsis* root. (a)** Root cross section diagram. Epidermal cells touching two cortical cells are in the H position, other epidermal cells are in the N position. **(b)** The transcription factor network in differentiated wild type epidermal cells. The epidermal cell touching the cortical cleft is in the H position, the other epidermal cell is in the N position. Black arrows represent translational promotion, black barbed arrows represent translational repression, and white arrows represent complex formation. Movement of proteins between epidermal cells is shown as grey arrows, dashed for diffusion and solid for SCM import of CPC. The cortical signal is represented as an arrow approaching the cortical signal (CS) receptor. Oval shapes represent mRNA and rectangles represent protein. For simplification of diagram, where transcriptional promotion arrows point straight to the protein, mRNA transcription and subsequent protein translation is inferred.

Central to the transcription factor network is a transcription factor complex comprising of the MYB transcription factors WEREWOLF (WER) and MYB23, the basic helix-loop-helix transcription factors GLABRA3 (GL3) and ENHANCER OF GLABRA3 (EGL3), and the WD-repeat protein TRANSPARENT TESTA GLABRA1 (TTG1)[2,4–6]. This WER/MYB23 complex specifies atrichoblast cell fate by promoting transcription of *GLABRA2* (*GL2*)[7–9], a homeodomain-Zip transcription factor responsible for suppressing root hair formation. Consistent with this role, WER, MYB23 proteins accumulate preferentially in N-position cells[4,5]. The WER/MYB23 complex promotes *MYB23* transcription[5], which in turn binds GL3, EGL3 to reinforce WER/MYB23 complex formation, creating a MYB23 positive feedback loop. Despite the similarities between *WER* and *MYB23*, the WER/MYB23 complex does not regulate *WER* transcription[5,10]. The WER/MYB23 complex also represses *GL3, EGL3* transcription[11,5] forming a negative feedback loop. Finally, it promotes transcription of *CAPRICE* (*CPC*)[12,13], and CPC competes with WER/MYB23 for GL3, EGL3 binding to form the CPC complex, preventing WER/MYB23 complex formation[6,9,14,15], in a second negative feedback loop.

Early studies suggested that *WER* transcription was negatively regulated by a cortical signal via the leucine-rich repeat receptor-like kinase SCRAMBLED (SCM)[16]. The role of SCM has since been revised, and SCM is now thought to function primarily in epidermal cell-cell signalling[17]. Nonetheless, as trichoblast cells form in files over cortical clefts and the WER/MYB23 complex promotes atrichoblast cell fate, *WER* is still hypothesised to be negatively regulated by an unidentified cortical signal. Positional information propagates through the epidermis via epidermal cell-cell signalling; CPC moves between epidermal cells via diffusion[18] and SCM-mediated import[17], GL3 moves between epidermal cells via diffusion[10,11,19].

Extensive characterisation of individual components and interactions have provided insight into *Arabidopsis* root epidermal patterning, none-the-less several fundamental questions remain unresolved. It is unclear whether *WER* transcription is subject to additional regulation beyond the cortical signal. A previous modelling study showed that additional regulation of *WER* transcription was necessary to reproduce epidermal patterning data, and suggested repression by CPC[10]. However, this study predated the discovery of MYB23[5] and modelled SCM in its previously accepted role as a cortical signal receptor[20]. In contrast, an experimental study hypothesised that competitive binding of GL3/EGL3 by WER and CPC alone may be sufficient to enable patterning[9]. As such, the extent of WER transcriptional regulation remains unresolved. SCM was initially thought to be a cortical signal receptor because trichoblast fate became partially decoupled from cortical clefts in *scm* mutants[20]. The redefinition of SCM's role as a CPC importer[17], raises the question, in *scm* mutants, is loss of CPC import sufficient to account for *scm* mutant data? It is not known what, if any, functional role the restriction of EGL3 movement has in epidermal patterning. Multiple binding sites have been reported for three reactions within the transcription factor network; WER binding sites on the *MYB23*[5], and *CPC*[12,13] promotors, and WER/MYB23 complex formation involving multiple GL3, EGL3, WER, MYB23 proteins[9]. While cooperativity and oligomerisation are expected to impact reaction dynamics, their contribution to trichoblast patterning has not been established.

To address these open questions and understand how components of the *Arabidopsis* root epidermal transcription factor network function together to generate robust patterning, mathematical modelling was used. A model was built which integrated the vast body of work describing network components and regulatory interactions. The model was used to assess the sufficiency of current knowledge to reproduce epidermal patterning, and perform mechanistic investigations designed to examine the essential nature and functional limits of network interactions. Specifically, the model was used to (i) propose and evaluate a new regulatory interaction, supported by promotor sequence analysis; (ii) test the functional contribution of SCM in its newly defined role as a CPC importer; (iii) examine the significance of asymmetric GL3 and EGL3 mobility; (iv) determine how cooperativity and oligomerisation contribute to patterning; and (v) characterise network behaviour in terms of fundamental patterning mechanisms. Together, these analyses provide an in-depth, mechanistic understanding of the transcription factor network governing *Arabidopsis* root hair patterning.

# Methods

## Model Description.

The purpose of this study was to understand the sufficiency of our current knowledge to reproduce *Arabidopsis* root epidermal patterning, and to gain an intuitive understanding of the patterning mechanisms within this complex transcription factor network. A reaction-diffusion model was created encompassing all regulatory interactions reported to date. Differential equation terms were derived using either mass action kinetics or the Hill equation[21]. To ensure generality and scalability, the model was formulated using arbitrary units (AU) for concentration, space and time.

The model was solved in a uniformly discretised one-dimensional space with periodic boundary conditions, representing a closed ring of epidermal cells. Each compartment corresponds to one epidermal cell, length $1$ AU. Protein and mRNA concentrations within each compartment are modelled as well mixed. Some model proteins can move between compartments, either passively via diffusion or actively via a transport reaction. Diffusion between cells was implemented using the standard finite difference scheme for the Laplacian operator, $\Delta$.

Epidermal cells overlying a cortical cleft are thought to receive positional signalling[16]. Cortical cleft- epidermal cell relationships were positionally mapped for twenty-two root cross-sections, yielding nineteen distinct topologies (Supplementary File S1). The model was either solved in the nineteen cross-section topologies or in a representative topology;

H-N-H-N-H-N-H-N-N-H-N-N-H-H-N-N-H-N-N,

where 'H' and 'N', represent an epidermal cell in the H or N-position, respectively.

Models were solved numerically in MATLAB, using bespoke solvers. All code is available on GitHub [22].

**Notation.**

Let lower case letters denote mRNA concentrations and upper-case letters denote protein concentrations. $i$ denotes the epidermal cell index. Thus, the concentration of protein $X$ in cell $i$ is denoted $X_i$, and $x_i$ denotes the concentration of mRNA $x$ in cell $i$.

Two classes of protein complexes were modelled: WER/MYB23 complexes, referred to as activator complexes, denoted $A$, and inhibitory CPC complexes, $I$. The CPC complex was modelled as a proxy for the CPC/TRY complex, as TRY is a homologue of CPC[15] and is also positively regulated by the WER/MYB23 complex, via GL2[23]. Although all complexes contain TTG1[2,6], TTG1 was not modelled explicitly due to the lack of regulatory data and was assumed not to be a limiting factor for complex formation.

***GL3, EGL3.***

*GL3* and EGL3, $g_i$ and $e_i$, are produced at basal rates $b_g$ and $b_e$, degraded at rates $d_g$ and $d_e$, and repressed by the total WER/MYB23 complex concentration within the cell, $A_{T_i}$, Equation (18)[11,5]. Repression was modelled as a degradation term with degradation rates $r_g$ and $r_e$, respectively.

$$\frac{d}{dt} g_i = b_g - r_g A_{T_i} g_i - d_g g_i \quad (1)$$

$$\frac{d}{dt} e_i = b_e - r_e A_{T_i} e_i - d_e e_i \quad (2)$$

**GL3, EGL3.**

GL3 and EGL3 proteins, denoted $G_i$and $E_i$, are translated at rates $q_G$ and $q_E$, degraded at rates $d_G$ and $d_E$, and diffuse between neighbouring epidermal cells with diffusion coefficients $D_G$ and $D_E$ [11,24].

GL3, EGL3 are members of the WER/MYB23 and CPC complexes[6,9]. The WER/MYB23 complex contains at least two GL3/EGL3 and two WER/MYB23 proteins[9]. To test the effect of complex oligomerisation, the number of proteins sequestered to WER/MYB23 complex formation was chosen as $n_1 \in \{1,2,3,4\}$, for GL3, EGL3 and $n_2 \in \{1,2,3,4\}$ for WER, MYB23. Upon complex dissociation, these proteins were returned to their respective pools. Because of lack of evidence to the contrary, one GL3, EGL3 protein is sequestered to CPC complex formation and returned on CPC complex deformation.

WER, MYB23, and CPC protein concentrations in cell $i$ are denoted $W_i$, $M_i$, and $C_i$, respectively. Complex concentrations are denoted $A_{GW_i}$, $A_{GM_i}$, $A_{EW_i}$, and $A_{EM_i}$ for the various combinations of the WER/MYB23 complex, and $I_{GC_i}$ and $I_{EC_i}$ denote the CPC complexes. Reaction rates $k_{11}$ to $k_{18}$ describe WER/MYB23 complex formation and deformation, and $k_{21}$ to $k_{24}$ describe CPC complex formation and deformation.

$$\frac{\partial}{\partial t} G_i = q_G g_i - n_1 G_i^{n_1}\left(k_{11} W_i^{n_2} + k_{15} M_i^{n_2}\right) + n_1\left(k_{12} A_{GW_i} + k_{16} A_{GM_i}\right) - k_{21} G_i C_i + k_{22} I_{GC_i} - d_G G_i + D_G \Delta G \quad (3)$$

$$\frac{\partial}{\partial t} E_i = q_E e_i - n_1 E_i^{n_1}\left(k_{13} W_i^{n_2} + k_{17} M_i^{n_2}\right) + n_1\left(k_{14} A_{EW_i} + k_{18} A_{EM_i}\right) - k_{23} E_i C_i + k_{24} I_{EC_i} - d_E E_i + D_E \Delta E \quad (4)$$

***WER.***

*WER* concentration, $w_i$, is increased by basal transcription, rate $b_w$, and degraded at rate $d_w$. *WER* transcription is repressed by a cortical signal[3], represented by $X_i \in \{0,1\}$, where $X_i = 1$ if cell $i$ overlies a cortical cleft. This is received by the epidermal cells via an unidentified cortical signal receptor, denoted $R_{X_i}$.

A previous modelling study suggested that *WER* transcription may also be repressed by CPC[10]. Also investigated in this study was the repression of *WER* transcription by the total concentration of CPC complex, denoted $I_{T_i}$, Equation (19). *WER* transcriptional repression was modelled as effective degradation with rates $r_{w11}$, $r_{w12}$, and $r_{w21}$. $r_{w11} r_{w12} = 0$ as only unbound CPC *or* CPC complex *or* neither repress *WER* transcription in any one model variant.

$$\frac{d}{dt} w_i = b_w - X_i\left(r_{w21} R_{X_i}\right) w_i - \left(r_{w11} C_i + r_{w12} I_{T_i} + d_w\right) w_i \quad (5)$$

### *MYB23.*

*MYB23*, denoted $m_i$, is degraded at rate $d_m$ and transcription is promoted by total WER/MYB23 complex, $A_{T_i}$. There are four WER binding sites on the *MYB23* promoter[5]. Thus, the number of WER/MYB23 complexes needed for *MYB23* transcription, $n_3 \in \{1,2,3,4\}$, was investigated.

A mass action kinetics term was initially used to describe promotion of *MYB23* transcription, $p_m A_{T_i}^{n_3}$, where $p_m$ is the transcription rate. It was found that using mass action kinetics for transcriptional promotion resulted in an unlikely *MYB23* concentration amplification for $n_3 > 1$. Thus, the Hill equation[21], rather than mass action kinetics, was used to describe promotion of *MYB23* transcription. The Hill coefficient is a measure of cooperativity between binding sites, with a maximum value equalling the number of available binding sites[25]. Thus, the Hill coefficient for *MYB23* transcription was set to $n_3 \in \{1,2,3,4\}$. The maximum rate of *MYB23* transcription was denoted $p_m$, and $K_m$ represents the dissociation constant, as derived by mass action kinetics.

$$\frac{d}{dt} m_i = p_m \frac{A_{T_i}^{n_3}}{K_m + A_{T_i}^{n_3}} - d_m m_i \tag{6}$$

### WER, MYB23.

WER and MYB23 are translated with rates $q_W$ and $q_M$, and degraded with rates $d_W$ and $d_M$, respectively. $n_2 \in \{1,2,3,4\}$ WER and MYB23 proteins are sequestered during the formation of the WER/MYB23 complex and gained when the WER/MYB23 complex disassociates[9].

$$\frac{d}{dt} W_i = q_W w_i - n_2 W_i^{n_2}\left(k_{11} G_i^{n_1} + k_{13} E_i^{n_1}\right) + n_2\left(k_{12} A_{GW_i} + k_{14} A_{EW_i}\right) - d_W W_i \tag{7}$$

$$\frac{d}{dt} M_i = q_M m_i - n_2 M_i^{n_2}\left(k_{15} G_i^{n_1} + k_{17} E_i^{n_1}\right) + n_2\left(k_{16} A_{GM_i} + k_{18} A_{EM_i}\right) - d_M M_i \tag{8}$$

### *CPC.*

*CPC*, denoted $c_i$, is degraded at rate $d_c$, and transcription is promoted by total WER/MYB23 complex [5,12,13]. Three distinct WER binding sites have been found on the *CPC* promotor, with all shown to be required for proper gene expression [12,13]. Thus, the number of WER/MYB23 complexes required for effective *CPC* transcription, was chosen from the set $n_4 \in \{1,2,3\}$. To prevent the biologically unrealistic amplification of *CPC* concentrations associated with mass action kinetics, transcriptional activation was modelled using a Hill function with Hill coefficient $n_4$, maximum transcription rate $p_c$, and dissociation constant $K_c$.

$$\frac{d}{dt} c_i = p_c \frac{A_{T_i}^{n_4}}{K_c + A_{T_i}^{n_4}} - d_c c_i \tag{9}$$

### CPC.

CPC is translated with rate $q_C$, degrades with rate $d_C$, and participates in CPC complex formation [6,9] with association and dissociation rates $k_{21}$ to $k_{24}$. CPC diffuses between epidermal cells [18] with diffusion coefficient $D_C$, and is actively imported via SCM, $S_i$, [17] with proportionality constant $\widetilde{D}_C$. The function describing SCM-mediated CPC import, $f(C,S)$, is given in Equation (11), where $i \pm 1$ index neighbouring epidermal cells. As SCM has been shown to localise in H-position cells [26], $S_i = 1$ when the epidermal cell lies over a cortical cleft and $0$ otherwise.

$$\frac{\partial}{\partial t} C_i = q_C c_i + k_{22} I_{GC_i} + k_{24} I_{EC_i} - (k_{21} G_i C_i + k_{23} E_i C_i) - d_C C_i + \widetilde{D}_C\, f(C,S) + D_C \Delta C \tag{10}$$

$$f(C,S) = \left(S_i(C_{i+1} + C_{i-1}) - C_i(S_{i+1} + S_{i-1})\right) \tag{11}$$

### WER/MYB23 complex.

$n_1 \in \{1,2,3,4\}$ GL3, EGL3 proteins and $n_2 \in \{1,2,3,4\}$ WER and MYB23 proteins are sequestered during WER/MYB23 complex formation and returned to the respective pools on complex deformation [9] with association and dissociation rates $k_{11}$ to $k_{18}$.

$$\frac{d}{dt} A_{GW_i} = k_{11} G_i^{n_1} W_i^{n_2} - k_{12} A_{GW_i} \tag{12}$$

$$\frac{d}{dt}A_{EW_i} = k_{13}E_i^{n_1}W_i^{n_2} - k_{14}A_{EW_i} \quad (13)$$

$$\frac{d}{dt}A_{GM_i} = k_{15}G_i^{n_1}M_i^{n_2} - k_{16}A_{GM_i} \quad (14)$$

$$\frac{d}{dt}A_{EM_i} = k_{17}E_i^{n_1}M_i^{n_2} - k_{18}A_{EM_i} \quad (15)$$

**CPC complex.**

GL3, EGL3 and CPC proteins are sequestered during CPC complex formation and returned to the respective pools on CPC complex deformation [6,9] with association and dissociation rates $k_{21}$ to $k_{24}$.

$$\frac{d}{dt}I_{GC_i} = k_{21}G_iC_i - k_{22}I_{GC_i} \quad (16)$$

$$\frac{d}{dt}I_{EC_i} = k_{23}E_iC_i - k_{24}I_{EC_i} \quad (17)$$

**Total WER/MYB23 and CPC complexes.**

As MYB23 can functionally replace WER [5] and GL3, EGL3 are functionally similar [6,9,11], total complex concentrations were defined as:

$$A_{T_i} = A_{GW_i} + A_{EW_i} + A_{GM_i} + A_{EM_i} \quad (18)$$

$$I_{T_i} = I_{GC_i} + I_{EC_i} \quad (19)$$

## Parameter Search.

There are a lack of data measuring the rates of reactions within the transcription factor network controlling *Arabidopsis* root epidermal patterning. A random search approach was used to explore the ability of the model to reproduce wild type (WT) and *scm* mutant root imaging data.

Thirty-eight core parameters were defined, Table 1. A core parameter set consisted of one value for each core parameter. $20{,}000$ core parameter sets were chosen uniformly at random from the arbitrary interval $(0{,}10]$. Core parameter sets were indexed $1$ to $20{,}000$ and fixed, i.e. core parameters defined reaction rates that remained unchanged across all modelling investigations. The values of eleven additional parameters were set depending on the mechanistic investigation being undertaken, Table 2. Within each parameter set, parameter values could vary in size by up to six orders of magnitude, whilst individual parameters varied across parameter sets by between three and six orders of magnitude (Supplementary File S2).

**Table 1. Core parameters.** Core parameter names and descriptions. $20{,}000$ core parameter sets were chosen uniformly at random from the interval $(0,10]$. $j$ is the index used to identify each core parameter during univariate sensitivity analysis.

| $j$ | | **WER/MYB23 complex formation and deformation** |
|---|---|---|
| 1 | $k_{11}$ | $A_{GW}$ association |
| 2 | $k_{12}$ | $A_{GW}$ disassociation |
| 3 | $k_{13}$ | $A_{EW}$ association |
| 4 | $k_{14}$ | $A_{EW}$ disassociation |
| 5 | $k_{15}$ | $A_{GM}$ association |
| 6 | $k_{16}$ | $A_{GM}$ disassociation |
| 7 | $k_{17}$ | $A_{EM}$ association |
| 8 | $k_{18}$ | $A_{EM}$ disassociation |
| | | **CPC complex formation and deformation** |
| 9 | $k_{21}$ | $I_{GC}$ association |
| 10 | $k_{22}$ | $I_{GC}$ disassociation |
| 11 | $k_{23}$ | $I_{EC}$ association |
| 12 | $k_{24}$ | $I_{EC}$ disassociation |
| | | **basal degradation** |
| 18 | $d_G$ | GL3 protein |
| 19 | $d_E$ | EGL3 protein |
| 20 | $d_C$ | CPC protein |
| 21 | $d_W$ | WER protein |
| 22 | $d_M$ | MYB23 protein |
| 23 | $d_g$ | *GL3* mRNA |
| 24 | $d_e$ | *EGL3* mRNA |
| 25 | $d_c$ | *CPC* mRNA |
| 26 | $d_w$ | *WER* mRNA |
| 27 | $d_m$ | *MYB23* mRNA |

| $j$ | | **basal transcription** |
|---|---|---|
| 15 | $b_g$ | *GL3* mRNA |
| 16 | $b_e$ | *EGL3* mRNA |
| 17 | $b_w$ | *WER* mRNA |
| | | **maximum transcription rate** |
| 33 | $p_c$ | *CPC* mRNA |
| 34 | $p_m$ | *MYB23* mRNA |
| | | **disassociation constant** |
| 37 | $K_m$ | WER on *MYB23* transcript |
| 38 | $K_c$ | WER on *CPC* transcript |
| | | **transcriptional repression** |
| 35 | $r_g$ | *GL3* by WER/MYB23 complex |
| 36 | $r_e$ | *EGL3* by WER/MYB23 complex |
| | | **translation** |
| 28 | $q_G$ | GL3 protein |
| 29 | $q_E$ | EGL3 protein |
| 30 | $q_C$ | CPC protein |
| 31 | $q_W$ | WER protein |
| 32 | $q_M$ | MYB23 protein |
| | | **protein movement** |
| 13 | $D_G$ | GL3 diffusion |
| 14 | $D_C$ | CPC diffusion |

**Table 2. Parameters changed during mechanistic investigations.** Parameter names, descriptions and intervals from which parameter values were chosen. Parameters were set depending on the mechanistic investigation being undertaken. Descriptions of mechanistic investigations are given in the sections indicated at the top of each parameter column. For each of the $20,000$ core parameter sets, the parameters $n_1$, $n_2$, $n_3$ and $n_4$ were chosen either uniformly at random from the set shown, or as described in the indicated tables. When parameter intervals are given the parameter was chosen uniformly at random from the interval, for each of the $20,000$ core parameter sets.

| | | **§ 3.1** | **§ 3.5** | **§ 3.6** |
|---|---|---|---|---|
| | **number of proteins in WER/MYB23 complex** | | | |
| $n_1$ | GL3, EGL3 | $\{2,3,4\}$ | Table 7. | $\{2,3,4\}$ |
| $n_2$ | WER, MYB23 | $\{2,3,4\}$ | Table 7. | $\{2,3,4\}$ |
| | **number of binding sites on transcript** | | | |
| $n_3$ | WER on *MYB23* | $\{2,3,4\}$ | Table 7. | $\{2,3,4\}$ |
| $n_4$ | WER on *CPC* | $\{2,3\}$ | Table 7. | $\{2,3\}$ |
| | **WER transcriptional repression** | | | |
| $r_{w11}$ | by CPC | Table 5. | Table 5. | Table 5. |
| $r_{w12}$ | by CPC complex | Table 5. | Table 5. | Table 5. |
| | **EGL3 movement** | | | |
| $D_E$ | diffusion | $0$ | $0$ | $(0,10]$ |
| | **SCM function** | | | |
| $S$ | SCM | $1$, SCM+<br>$0$, SCM- | | |
| $\widetilde{D}_C$ | CPC import via SCM | $(0,10]$ | | |
| | **Cortical signal input** | | | |
| $R_X$ | unknown receptor | $(0,0.25]$ | | |
| $r_{w21}$ | *WER* repression via $R_X$ | $(0,10]$ | | |

## Experimental datasets.

The WT data in Table 3 was generated by analysing twenty-two root cross-section images containing a GL2 reporter, Fig. 2a. Published *scm* mutant data was available in table form[16,9,27,17,28], however, reported standard deviations were much smaller than those calculated using model solutions and from analysing published images. To ensure consistent statistical treatment, *scm* mutant data was calculated from eleven figures within six publications[20,16,9,27,17,28], Table 3. The means calculated using the published imaging data were comparable to the published tabulated means and their standard deviations comparable to those calculated using model solutions.

## Model evaluation against biological data.

During parameter searches, models were solved within the representative cross-section topology. Steady state solutions of models which included SCM, referred to as SCM+ models, were compared to wild type (WT) data. Steady state solutions of models lacking SCM, referred to as SCM- models, were compared to *scm* mutant data. Biological data regarding the location of transcript and proteins were imaging data. Imaging data for protein localisation indicates total protein concentrations within a cell, without distinguishing between free or bound species. Accordingly, model outputs were compared using total concentrations.

In WT, trichoblast cells generally form in single files over the cortical cleft. Thus, it can be assumed that transcription factor network components that follow this pattern are co-localised in trichoblast cells. A similar argument can be made for atrichoblast transcription factor network components. Therefore, for SCM+ model solutions, *in-silico* trichoblast cells were defined as having a steady state solution with high concentrations of total CPC[29], *GL3*, *EGL3,* and total EGL3[6,24], and low concentrations of *CPC*[30]*,* total GL3[11], *WER*, total WER[4], *MYB* and total MYB[5], when compared to atrichoblast cells. *In-silico* atrichoblast cells were defined conversely, Fig. 1b. A concentration difference threshold of 10% was required to classify differences as significant. The positions of ‘well defined’ *in-silico* trichoblast, atrichoblast cells were then compared to WT data, Table 3.

In *scm* mutant roots, as cell fate is decoupled from the cortical cleft, and localisation data is not available for many of the transcription factor network components, co-localisation cannot be assumed. Therefore, for SCM- model solutions, the *in-silico* WER/MYB complex (used as a *GL2* proxy), total WER and *CPC* were compared to *scm* mutant data, Table 3, as these are the only available markers in the *scm* mutant[20,16,9,27,17,28].

A parameter set was labelled successful if the SCM+ model steady state solutions reproduced WT data and SCM- model steady state solutions reproduced *scm* mutant data, Table 3.

**Table 3. WT and *scm* mutant cell type and position data.** Mean percentage and standard deviations of each epidermal cell type in the H and N positions. Full data sets in Supplementary Files S1 and S2.

| | **H position** | | **N position** | |
|---|---|---|---|---|
| | **trichoblast** | **atrichoblast** | **trichoblast** | **atrichoblast** |
| **WT** | $93.7 \pm 8.4$ | $6.3 \pm 8.4$ | $1.2 \pm 3.0$ | $98.8 \pm 3.0$ |
| ***scm*** | $61.8 \pm 12.9$ | $38.2 \pm 12.9$ | $22.4 \pm 15.2$ | $77.6 \pm 15.2$ |

## Parameter Search Protocol.

For each mechanistic combination investigated, each parameter set $P_i$, $i = 1, \dots, 20{,}000$, where $P_i$ is a list of forty-nine parameter values, one for each of the parameters shown in Tables 1 and 2, was tested using a three step parameter test protocol as follows.

**Step 1: WT Patterning in the SCM+ model.**

The SCM+ model was solved with zero initial conditions. If the steady state solution was such that all *in-silico* trichoblast cells occupied H-positions and all *in-silico* atrichoblast cells occupied N-positions, the parameter set was used in step 2; otherwise, it was discarded.

**Step 2: *scm* mutant patterning in the SCM- model.**

The SCM- model was solved using random initial conditions. As *WER* transcription is influenced by the cortical signal, the steady state solution of the SCM- model, solved with uniform initial conditions of any value, would tend to pattern in line with the cortical clefts. Thus, random initial conditions were used when solving the SCM- model.

Each model variable was assigned an initial condition, $C_0$, chosen uniformly at random from the interval $(0, C_{0_{max}}]$, $C_{0_{max}} \leq 1$. To fit SCM- model solutions to *scm* mutant data, the cortical signal receptor concentration, $R_X \in (0, 0.25]$, and the magnitude of the random initial conditions, $C_{0_{max}} \in (0,1]$, were incrementally searched. For each $R_X$, $C_{0_{max}}$ combination, twenty independent random initial conditions were generated and used to solve the SCM- model.

The twenty SCM- steady state solutions were used to calculate the mean percentage and standard deviations of trichoblast cells in H-positions and atrichoblast cells in N-positions. These distributions were then compared to *scm* mutant data using Welch's two-sided t-test (95% confidence interval). If the absolute value of the t-statistic was less than the critical t-value, then *in-silico* and *in-planta* data were considered not significantly different and parameter set was labelled as successful and used in Step 3. If no $R_X$, $C_{0_{max}}$ combination could be found which produced t-statistics within the critical bounds, the parameter set was discarded.

**Step 3: Robustness of WT patterning.**

The SCM+ model was solved again using the fitted $R_X$ and $C_{0_{max}}$ values identified in Step 2, with twenty random initial conditions to check that WT data could still be reproduced with the full parameterisation.

The set of twenty SCM+ solutions were then compared to WT data using Welch's two-sided t-test (95% confidence interval). If the resulting t-statistic had absolute value less that the critical t-value then *in-silico* and *in-planta* data were considered not significantly different and parameter set $P_i$ was labelled as a successful parameter set, otherwise $P_i$ was considered unsuccessful and discarded.

Parameter values for the cortical signal receptor used in each step are summarised in Table 4. All parameter sets can be found on GitHub [22].

**Table 4. SCM and cortical signal receptor parameter values for the three parameter test steps.**

| | initial conditions | SCM, $S$ | receptor, $R_X$ |
|---|---|---|---|
| Step 1: SCM+ | zero | $1$ | $1$ |
| Step 2: fit SCM- | random $= (0,1]$ | $0$ | $(0,0.25]$ |
| Step 3: SCM+ | random $= (0,1]$ | $1$ | $(0,0.25]$ |

## WER transcriptional regulation.

Alternative regulatory mechanisms governing WER transcription were investigated by modifying Equation (5). The parameter values used for each investigation are given in Table 5, and the resulting functional forms of Equation (5) are shown in Table 6.

For each WER regulation mechanism, model outputs were compared to SCM+ and SCM- steady state solutions. Qualitative comparisons were also performed by comparing imaging data with steady state solutions, for parameter sets and topologies chosen at random.

Mutant models were generated using the SCM+ model by equating the differential equation describing the mutated component(s) to zero. For example, in the *wer* mutant model, Equation (5) was changed to $\mathrm{d}w_i/\mathrm{dt} = 0$. All mutant models were solved in the representative cross-section topology (Supplementary File S1) for twenty sets of random initial conditions. Mutated component(s) were assigned zero initial concentration.

For tests against robustness to topology, for each parameter set and topology, the model was also solved with twenty random initial conditions, and the resulting distributions of *in-silico* means were compared to the *in-planta* distributions.

**Table 5. Mechanistic parameter intervals for *WER* transcriptional repression investigation.** CS stands for cortical signal. Other parameter intervals are shown in Tables 1 and 2.

| parameter descriptions | | parameter values for each mechanism | | |
|---|---|---|---|---|
| | **WER transcriptional repression** | **CS only** | **CS & CPC** | **CS & CPC complex** |
| $r_{w11}$ | by CPC | $0$ | $(0,10]$ | $0$ |
| $r_{w12}$ | by CPC complex | $0$ | $0$ | $(0,10]$ |

**Table 6. Forms of Equation (5) in the *WER* transcriptional repression investigation.**

| Equation (5) | |
|---|---|
| General form | $\frac{d}{dt}w = b_w - X(r_{w21}R_X)w - (r_{w11}C + r_{w12}I_T + d_w)w$ |
| **CS only** | $\frac{d}{dt}w = b_w - X(r_{w21}R_X)w - d_w w$ |
| **CS & CPC** | $\frac{d}{dt}w = b_w - X(r_{w21}R_X)w - (r_{w11}C + d_w)w$ |
| **CS & CPC complex** | $\frac{d}{dt}w = b_w - X(r_{w21}R_X)w - (r_{w12}I_T + d_w)w$ |

## Univariate Sensitivity Analysis.

Univariate sensitivity analysis was performed on successful parameter sets, $P_i$, within the representative cross-section topology. Each core parameter $P_{i,j}$, $j = \{1,2, \dots 38\}$, Table 1, was changed one by one. The mechanistic parameters within $P_i$, Table 2, were not changed during sensitivity analysis, as those parameters define specific biological mechanisms under scrutiny and to change them would be to change the biological mechanism.

For each core parameter $p_{i,j}$, the set of perturbed values, $p'_{i,j}$, was generated like so,

$$p'_{i,j} = 10^{x} p_{i,j}, \; x = \{-2, -1.75, -1.5, \dots, 2\}.$$

For each perturbed parameter, $p'_{i,j}$, the SCM+ and SCM- models were solved twenty times, each with different noisy initial conditions. The model solutions were then compared with experimental data, Table 3. The success or failure of the model solution, solved with parameter set $P_i$, containing the perturbed parameter $p'_{i,j}$, was recorded. An insensitive region for core parameter $p_{i,j}$ was defined as,

$$r_{i,j} = x_{max} - x_{min},$$

where $x_{min}$ was the minimum value of $x$ for which the model solved with parameter set $P_i$, containing the perturbed parameter $p'_{i,j}$ was a success, and $x_{max}$ was the maximum value of $x$.

Insensitive regions were calculated for all core parameters, $P_{i,j}$, for both SCM+ and SCM- models. The insensitive regions of SCM+ and SCM- models were then compared to understand if CPC import by SCM decreased the transcription factor network's sensitivity to reaction rate changes. The difference between insensitive regions was calculated for each core parameter:

$$\Delta r_{i,j} = {r_{i,j}}^{SCM+} - {r_{i,j}}^{SCM-}.$$

Where $r_{i,j}^{SCM+}$ is the insensitive region for SCM+ models and $r_{i,j}^{SCM-}$ for SCM- models. The resulting number $\Delta r_{i,j}$ defined how much less sensitive to changes in parameter $p_{i,j}$ the SCM+ model was, when compared to the SCM- model. If the difference for a given parameter $p_{i,j}$ was negative the SCM+ model, containing CPC import and diffusion, was more sensitive to changes in $p_{i,j}$ than the SCM- model, containing CPC diffusion only.

Sensitivity analysis code can be found on GitHub [22].

## Multiple binding sites.

The contribution of multiple binding sites to network behaviour was investigated by selectively removing binding site interactions from the model. Parameter intervals used for this analysis are shown in Table 7. For each configuration, one or more multiple binding site reactions were removed by setting the corresponding parameters to $1$. For each investigation, parameter searches were performed.

**Table 7. Mechanistic parameter intervals for the multiple binding site investigation.** Other parameter intervals are shown in Tables 1, 2 and 5. For one multiple binding site reaction removed, the name of the removed binding site reaction is shown. For two multiple binding site reactions removed, the name of the remaining binding site reaction is shown. $n_1$ and $n_2$ are parameters in the same multiple binding site reaction, WER/MYB23 complex formation.

| number of multiple binding site reactions removed | | 0 | 1 | | | 2 | | | 3 |
|---|---|---|---|---|---|---|---|---|---|
| parameter | location | | multiple binding site removed | | | multiple binding site remaining | | | |
| | | all | *CPC* | *MYB23* | WER/MYB23 complex | *CPC* | *MYB23* | WER/MYB23 complex | none |
| $n_1$ | GL3, EGL3 in complex | $\{2,3,4\}$ | $\{2,3,4\}$ | $\{2,3,4\}$ | $1$ | $1$ | $1$ | $\{2,3,4\}$ | $1$ |
| $n_2$ | WER, MYB23 in complex | $\{2,3,4\}$ | $\{2,3,4\}$ | $\{2,3,4\}$ | $1$ | $1$ | $1$ | $\{2,3,4\}$ | $1$ |
| $n_3$ | *MYB23* translation | $\{2,3,4\}$ | $\{2,3,4\}$ | $1$ | $\{2,3,4\}$ | $1$ | $\{2,3,4\}$ | $1$ | $1$ |
| $n_4$ | *CPC* translation | $\{2,3\}$ | $1$ | $\{2,3\}$ | $\{2,3\}$ | $\{2,3\}$ | $1$ | $1$ | $1$ |

## EGL3 restricted movement.

To investigate the effect of restricted EGL3 movement, models were solved with EGL3 diffusion enabled, $D_E > 0$, Table 2, for each *WER* regulation mechanism, Tables 5 and 6, for all $20{,}000$ core parameter sets.

The ten successful parameter sets found with $D_E = 0$, plus an extra parameter set found for $D_E > 0$, were analysed further. Each successful parameter set has twenty SCM+ solutions and twenty SCM- solutions recorded from the tests with twenty random initial conditions. This provided 220 steady state solutions for each of the SCM+ and SCM- models for analysis.

## Substrate depletion patterning mechanism

The necessity of substrate depletion for correct patterning was tested for each of the substrates in turn. WER/MYB23 complex inhibition of *GL3* or *EGL3* transcription removed by setting $r_g = 0$ or $r_e = 0$ in the SCM- no cortical signal model and a parameter search was performed.

## Experimental Methods.

### Plant Materials.

Wildtype *Arabidopsis* (Col-0) and *scm-2* mutants (SALK 086357) were obtained from the Nottingham *Arabidopsis* stock centre (NASC). Sterile seeds were sown onto 1% agar plates containing 0.5x Murashige and Skoog salts at pH 5.7. Plates were then stratified at 4°C for 48h before being moved to growth rooms set at 21°C and a 16-hour day length.

### Molecular cloning and plant transformation.

*Arabidopsis thaliana GLABRA2* (AT1G79840) genomic coding sequence was cloned from genomic DNA using primers designed according to the Greengate cloning system protocol[31]. The *GLABRA2* coding sequence was cloned with a silent site-directed mutagenesis (V438V) to remove an internal BsaI site. mTurquoise was cloned from an existing plasmid with primers which include a glycine-serine (Gly-Ser) linker (GSSGGGGSGGGGS) at the N-terminal end. The *GL2* promoter was cloned and described previously[32]. Restriction and ligation reactions were performed using BsaI and T4 DNA ligase (New England Biolabs, MA, USA). All entry plasmids were confirmed for presence of an insert by colony PCR and Sanger or Oxford Nanopore sequencing (Source Bioscience, Cambridge, UK). The entry vectors were used to generate an expression vector with GL2 fused with mTurquoise at the C-terminus under the control of *GL2* promoter (Supplementary File S3). Other components of the construct were derived from the Greengate kit described by Lampropoulos *et* al. (2013). This kit is available from Addgene (Kit #1000000036). The construct was assembled in the pGGZ001 Greengate destination vector (Supplementary File S3). Greengate assembly reactions were performed using NEB Golden Gate Cloning Kit (New England Biolabs, MA, USA). Final expression constructs were confirmed by colony PCR, restriction digest and Oxford Nanopore sequencing.

The expression vector was transformed into *Agrobacterium tumefaciens* by electroporation transformation in combination with pSOUP, which is necessary for proper replication of the pGREEN-based plasmid. *A. tumefaciens* colonies were selected using colony PCR for presence of the expression vector. These were cultured and then used to transform *Arabidopsis thaliana* plants via the floral dip method [33]. The construct was transformed into Columbia 0 background and *scm-2* (SALK_086357; insertion at *SCRAMBLED/STRUBBELLIG* AT1G11130; [20]). Transformant plants were selected for using hygromycin resistance.

### Imaging and image processing.

5-day old seedlings were fixed with 4% paraformaldehyde in phosphate-buffered saline pH7.4 and cleared according to the Clearsee method [34], staining for 1h with Direct Red 23 (Sigma Aldrich) during clearing as described by Ursache *et al*. (2018) [35]. After clearing the root tips were mounted on slides and imaged on a Leica SP8 confocal microscope. Direct red 23 was excited using a 561nm laser line and detected at 580-615nm. mTurquoise was excited using a 442nm laser line and detected at 450-518nm. Data was processed using FIJI (ImageJ) and cell identity counts based on presence or absence of GLABRA2 protein estimated by mTurquoise signal.

# Results

## The transcription factor network model can reproduce wild-type and *scm* mutant epidermal patterning if repression of *WER* is colocalised with the CPC complex.

*Arabidopsis* root epidermal patterning is controlled by a complex transcription factor network. It remains unclear whether current understanding of this network is sufficient to explain observed patterning. To address this, published regulatory interactions were integrated into one reaction network model (Methods). Mechanistic parameters (Table 2) were initially set to represent accepted mechanisms including SCM as a CPC importer, EGL3 restricted movement, cooperativity on *MYB23* and *CPC* transcription and oligomerisation within the WER/MYB23 complex.

*WER* transcription is known to be directly repressed by the cortical signal however, whether additional regulation is required remains unclear. A previous modelling study suggested that *WER* transcription may also be repressed by CPC[10], whereas an experimental study hypothesised that competitive binding for GL3/EGL3 by WER and CPC was sufficient to enable patterning. Both mechanisms were explicitly tested using the model (Tables 5 and 6).

No parameter sets reproduced biological data when *WER* transcription was regulated solely by the cortical signal, or by the cortical signal together with CPC. Parameter sets produced steady state solutions which were either homogeneous, patterned but did not pass the 10% threshold, have well defined cell types but trichoblast arrangement did not match the data in Table 3, or have poorly defined cell types (Supplementary File S4). No oscillators were found.

These results suggest that the competitive binding mechanism alone is insufficient to enable *Arabidopsis* root epidermal patterning. Additional regulation on *WER* by CPC was also inadequate, which is perhaps unsurprising as *CPC* transcription is promoted by, and therefore colocalised with, the WER/MYB23 complex. Thus, allowing unbound CPC to repress *WER* transcription, hindering the ability of the WER/MYB23 complex to become dominant.

It was hypothesised that successful patterning may require regulation of *WER* by a network component that dominates in trichoblast cells, and thus repression by the CPC complex was tested. *WER* transcriptional repression by the cortical signal and CPC complex enabled the model to reproduce both wild-type (WT) and *scm* mutant data, and ten successful parameter sets were identified (Supplementary File S4). Qualitative comparison between *in-planta GL2* expression and *in-silico* WER/MYB23 complex, a proxy for *GL2* in the model[7–9], showed close agreement for both WT and the *scm* mutant, Fig. 2a.

To further validate the model, additional mutant models were solved, namely, *wer*, *myb23*, *wer myb23*, *cpc*, *gl3*, *egl3*, and *gl3 egl3*. All mutant phenotypes were successfully reproduced apart from N-position cells in the *gl3* mutant, Fig. 2b (Supplementary File S4). This discrepancy likely reflects the difference in developmental stages between model solutions and experimental observations, as *in-silico* and *in-planta* comparison is reconciled when using mature ‘upper zone’ root data[6].

To evaluate model robustness to root geometry, the model was solved in nineteen topologies derived from root cross-section data. All ten parameter sets successfully reproduced WT data across all nineteen topologies, Fig. 2c (Supplementary File S4), demonstrating that the model is robust to variability in positional signalling. Successful parameter sets did not occupy a common area of parameter space and parameter relationships found held for many unsuccessful parameter sets (Supplementary File S4).

Together these results show that regulation of *WER* by the CPC complex, or a colocalised factor, is necessary to explain *Arabidopsis* root epidermal patterning. The model with *WER* repression by the CPC complex reproduced the nonuniform trichoblast spacing found in *Arabidopsis* roots as well as all mutant data tested. The model thus predicts the existence of a previously uncharacterised transcriptional regulation acting on *WER* in the root epidermis.

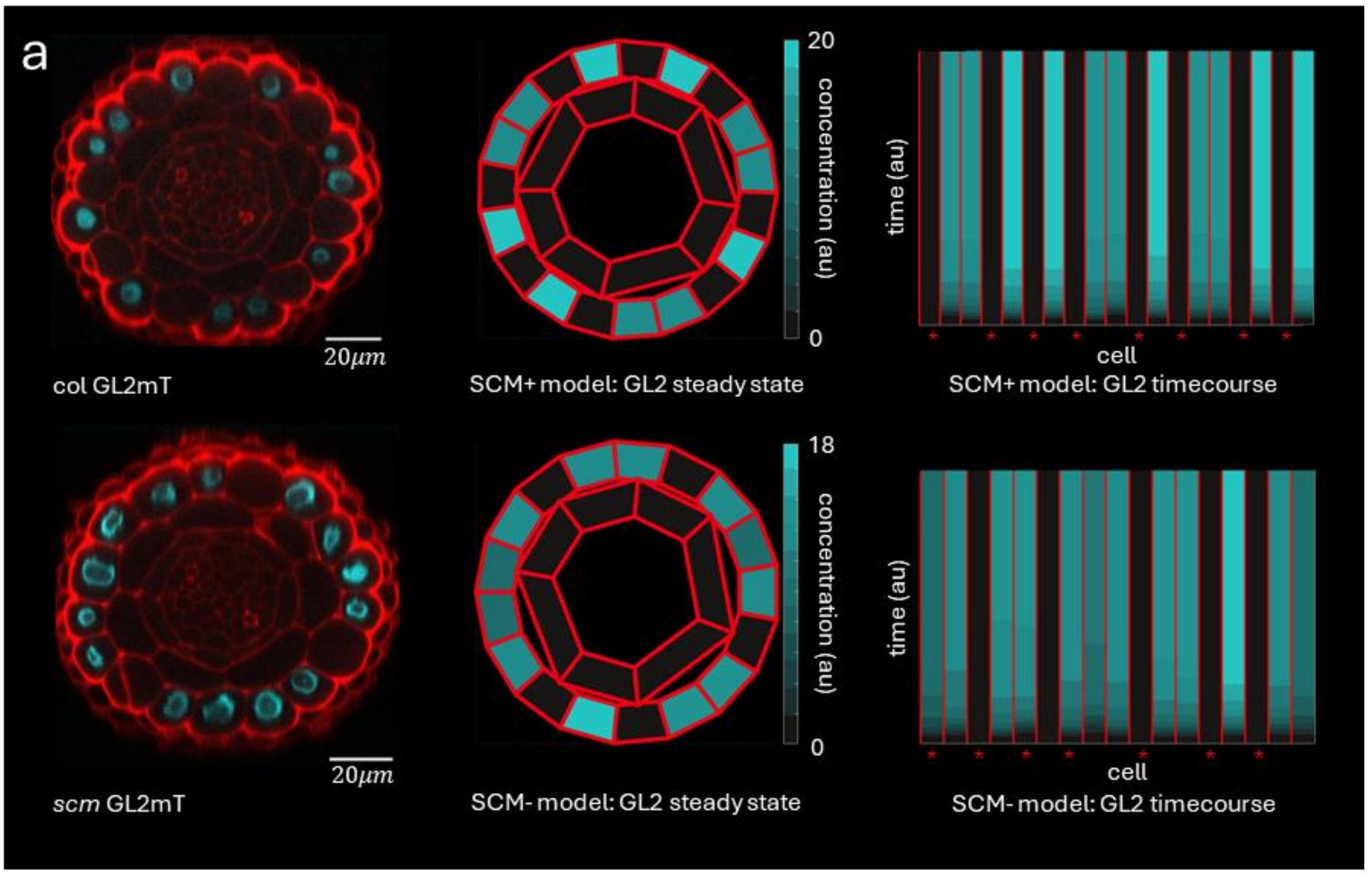

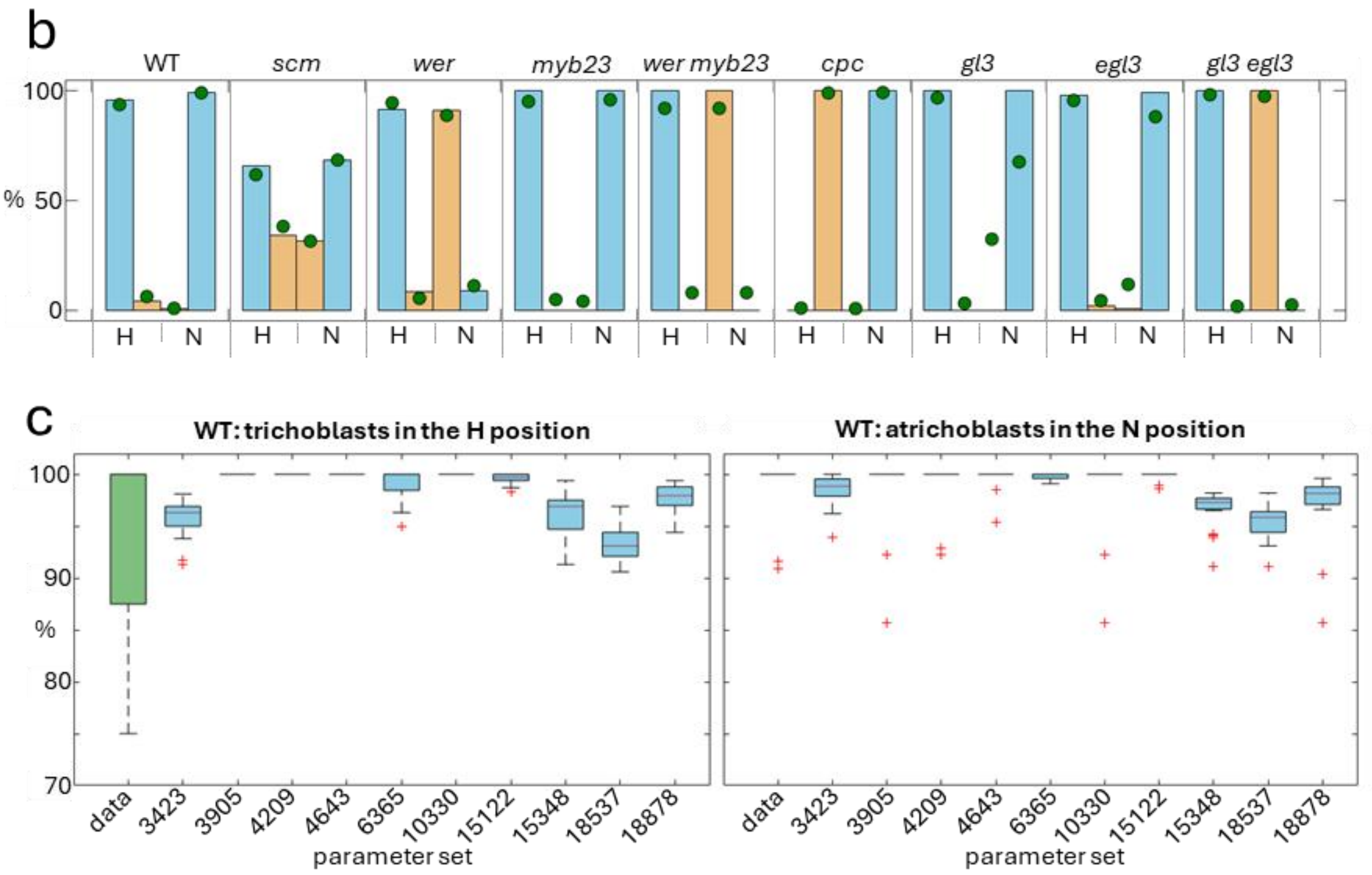

**Figure 2: Comparing in-planta and in-silico data for the model with *WER* transcriptional regulation by the CPC complex. (a)** Top row. Left, WT data. Middle, steady state solution of the SCM+ model, solved with parameter set $P_{15348}$. Right, model solution time course, stars on the horizontal axis show cortical cleft position. Bottom row shows *scm* mutant data and SCM- model results. SCM- model was also solved with parameter set $P_{15348}$. **(b)** Comparison between *in-planta* and *in-silico* data for WT and mutants. H represents H position, N the N position. Following the same pattern as in Table 3, the first data in each H or N position is percentage of trichoblast cells, the second data is percentage of atrichoblast cells. Green dots show mean percentages calculated from *in-planta* data, bars show mean *in-silico* percentages. Blue bars show *in-silico* means of trichoblast cells in the H position or atrichoblast cells in the N position calculated from 200 simulations, 20 repeats in each of the 10 successful parameter sets. Beige bars show the corresponding *in-silico* ectopic cell means. **(c)** Comparison between WT *in-planta* and SCM+ *in-silico* data in all root cross-section topologies. The first boxplot on each graph is the distribution of *in-planta* data, Table 3. Each of the other boxplots is the distribution of mean *in-silico* percentages for the nineteen topologies, for each parameter set.

## The CPC complex is a viable candidate for WER transcriptional regulation.

For the CPC complex to be a candidate for *WER* transcriptional repression, it must bind the *WER* promotor. CPC is an R3-type MYB, lacking the R2 repeat needed for sequence-specific DNA binding, and is therefore unlikely to bind the WER promoter directly[23]. However, the CPC complex contains GL3/EGL3, which are bHLH proteins capable of binding E-box (CANNTG) and G-box (CACGTG) motifs, with G-box motifs showing the strongest affinity in plants[36,37].

A targeted analysis of the *WER* (AT5G14750; MYB66) promotor (chr5:4,763,372–4,764,834, Araport11) [38], was performed in a 1,500 bp window upstream of the transcription start site. Multiple predicted GL3/EGL3 binding motifs were found including a G-box motif and several E-box variants (CACATG, CATGTT, CAGCTT)[37,39]. Notably, the specific region previously shown to be required for nonhair-cell-preferential WER expression[40], contains the high affinity G-box motif along with two of the non-canonical E-box motifs, with the third and two canonical E-box motifs clustered nearby, Fig. 3. While the presence of GL3/EGL3 binding sites on the *WER* promotor supports potential CPC complex occupancy, previous studies suggest that GL3/EGL3 binding sites alone may not be sufficient for stable complex binding and effective transcriptional regulation[6,41,42]. Therefore, other DNA-binding factors that could stabilise CPC-complex occupancy were considered.

The CPC complex component, TTG1 has been shown to physically interact with the WRKY protein TTG2 in *Arabidopsis* leaf[43]. TTG2 binds W-box motifs (TTGACC/TTGACT) in double-stranded DNA[44], thus raising the possibility that CPC-bound TTG1 could recruit TTG2 to stabilise promotor binding. It has been suggested that the role of TTG1 in a bHLH complex is to facilitate interactions with proteins that aid in the tethering of the complex to its target promotors[45]. Consistent with potential WRKY occupancy, two reverse-complement W-box matches (AGTCAA/GGTCAA) were found within the WER promotor region, Fig. 3. Genetic and transcriptomic evidence supports overlapping TTG2 and CPC function in root epidermal development[46], including the enhancement of the *cpc* mutant phenotype in the *cpc ttg2* double mutant[46,47].

Promotor sequence analysis coupled with published data, supports the CPC complex as a viable candidate for the predicted additional transcriptional regulation of *WER*, potentially via GL3/EGL3 binding of the *WER* promotor stabilised by TTG1-TTG2 interaction.

5’ TTGGATATGTGATATGCCCAAATAGATCATTCGAATTTTGTATGCAGAAAACTCTAATTCTTAG
TTATGGATGCAAAATATTAATTATGGATCTTATTATTATTGTTTGTCATATTGCTGATAAGATTACTGAT
ATAGTCAATTACTGAAAAGAAAATCATTTTGTTAATCTTGAATAAATAAGTCAGCAAAACAAGATGT
ACAATAGCAAAAATATCAACAAAATAAATTCTTTTTTTTCAAACAAAGTTAATATTTGCTTACAACA
CAGATTGGAAATTTCAAATATCCATCGCACATGCTATGTGACAGAAAAAAAATATGAACAATATAC
GTGATTATAAATTCAATAATCAATAGTAATAGTATCTATACGTAAATTAAAGATATTTGGTGGGGGCA
CAGAGATCAAGAAATTTTAATGATAAGAGAAGGAGCTTCTATTCTCTCTTTCTTTCTAACAATAACA
CTGTTTTTTTTTCCCCGTGAAGAATCCTAATCCCAAGAGAGTCAAAGTGTCAAACCATCTTCTAA
TTGAATATCTCCCCTAACTACAATTCCTAACTCTCATGCCACTAATCAATTACTTATTGTACATTTTT
CTTTTTTTAATTAGTAACCAAATTTACAAAATTAACTCTAGGAAATCGTTTAATCCAAGATTCCCGG
ACGTTGGTCCAAGGGTATCACTCTACAAAATAATTATAAAAAATGACGAGGCATTTCATTTTCTAA
TCAATGTATTTTCTGATAAAGGATGTCATATAAACTTGGTATCTCATAAATAAAGTATCTAAACTTCA
AAACGAACAACTACTATTACTTGATGAAAATAACAACACTATTATAATTAATGGTTAGTTGAGAAAC
AATACATAAAAAATTTGATGGAACTATGAGAGCCAGCCAGTTACTATCCTTCTCACCTTCCAATG
GGTTGCGGCAAGAACCTTCACAACCTCCTTTATATTTGCGATGTGTATCTAGTCTTTTGCAATGG
TCTCAAAACACTTTGGGTTTCATAAGTTTAACTATAATGGTGTCCCTGATATTTTCGGTCTAATATCT
CTTAAAAAAGAAAAATACTTATGATTCATTTCAATTAATCAAGGTTCAAGAAGATATATAAACACTA
GCCCTGACACATGAAACTTC**GATGCCGAAAAGCTCTAAGATCAAAGCCGAATCTTTTTAAAA**
**CATACACGTGATTTTTGTGTCTCCCAAGTCTTAACCGGTCCATGTTTTCATGTTT**TAGTTAGAAA
TCTAGATTAAGTCATTAAACTAATCCGTATCAGTAATTACCAGCTTGCATCTCAGAAGTCCATTATT
ATTTACATATCATCATCACATGCTAGACACAATCAATACCTTATGTCAATATCTAAAATAAGTCTATA
ATCATTAATACTTGTATTTATACCAATAGTATATCGTTTATTAAATATTATTCATACTTTATACATAAATAT
TCCACTAGGTTCTGAACTTGTAGTACTACTA 3’

**Figure 3: *WER* sequence.** 1,500 bp window upstream of the *WER* transcription start site. Canonical E-boxes, yellow. G-box, blue. E-box motifs, green. Reverse-complement W-boxes, red. The − 420 to − 346 bp region, bold type.

## Dysregulation of CPC movement is sufficient to reproduce *scm* mutant data.

SCM was previously hypothesised to function as a cortical signal receptor because its removal partially decouples trichoblasts from the cortical clefts[20]. With SCM redefined as a CPC importer[17], it was unclear whether dysregulation of CPC movement alone could account for the *scm* mutant phenotype. Successful parameter sets were required to reproduce *scm* mutant data when SCM, i.e. CPC import, was removed. The existence of successful parameter sets show that decoupling of trichoblast patterning from the cortical clefts can emerge without altering cortical signal perception, rather, decoupling can arise from dysregulation of CPC movement alone. This result provides a mechanistic explanation for why *scm* mutant patterning is not entirely random; dysregulation of CPC movement disrupts patterning but does so incompletely, trichoblast cells retain a weakened bias towards H positions due to the cortical signal.

## SCM import of CPC reduces epidermal patterning sensitivity to reaction rate changes.

CPC is predominantly transcribed in N positioned cells[12,13]. SCM import of CPC from N to H positioned cells likely aids patterning by reducing the CPC available to compete with WER and MYB23 for GL3, EGL3, allowing WER/MYB23 complex formation in N positioned cells[6,9]. If SCM import aids patterning in this way, one might expect that models with SCM import of CPC would be less sensitive to parameter changes than models with CPC diffusion alone.

Univariate sensitivity analysis was performed on the ten successful parameter sets, comparing SCM- models containing CPC diffusion alone to SCM+ models containing both CPC diffusion and import. For six of the ten parameter sets analysed, the SCM+ model was less sensitive than the SCM- model for all thirty-eight core parameters, while the remaining four contained a small subset of parameters for which the SCM+ model was more sensitive, Fig. 4a. Nonetheless, across all parameter sets, the total insensitive region for the SCM+ model was greater than the SCM- model, Fig. 4b, indicating that SCM import of CPC increases the transcription factor network’s robustness against reaction rate fluctuations.

Summing sensitivity differences for each parameter over all parameter sets, showed that the inclusion of CPC import expanded the range over which patterning could be maintained for all parameters, Fig. 4c. Notably, removal of CPC import disproportionately increased sensitivity to parameters governing EGL3 production and degradation, and CPC/EGL3 complex formation. In contrast, parameters associated with GL3 reactions were among the parameters least affected, grouped with those governing WER/MYB complex concentrations. Thus, the data suggest that SCM import of CPC preferentially supports formation of the CPC/EGL3 complex.

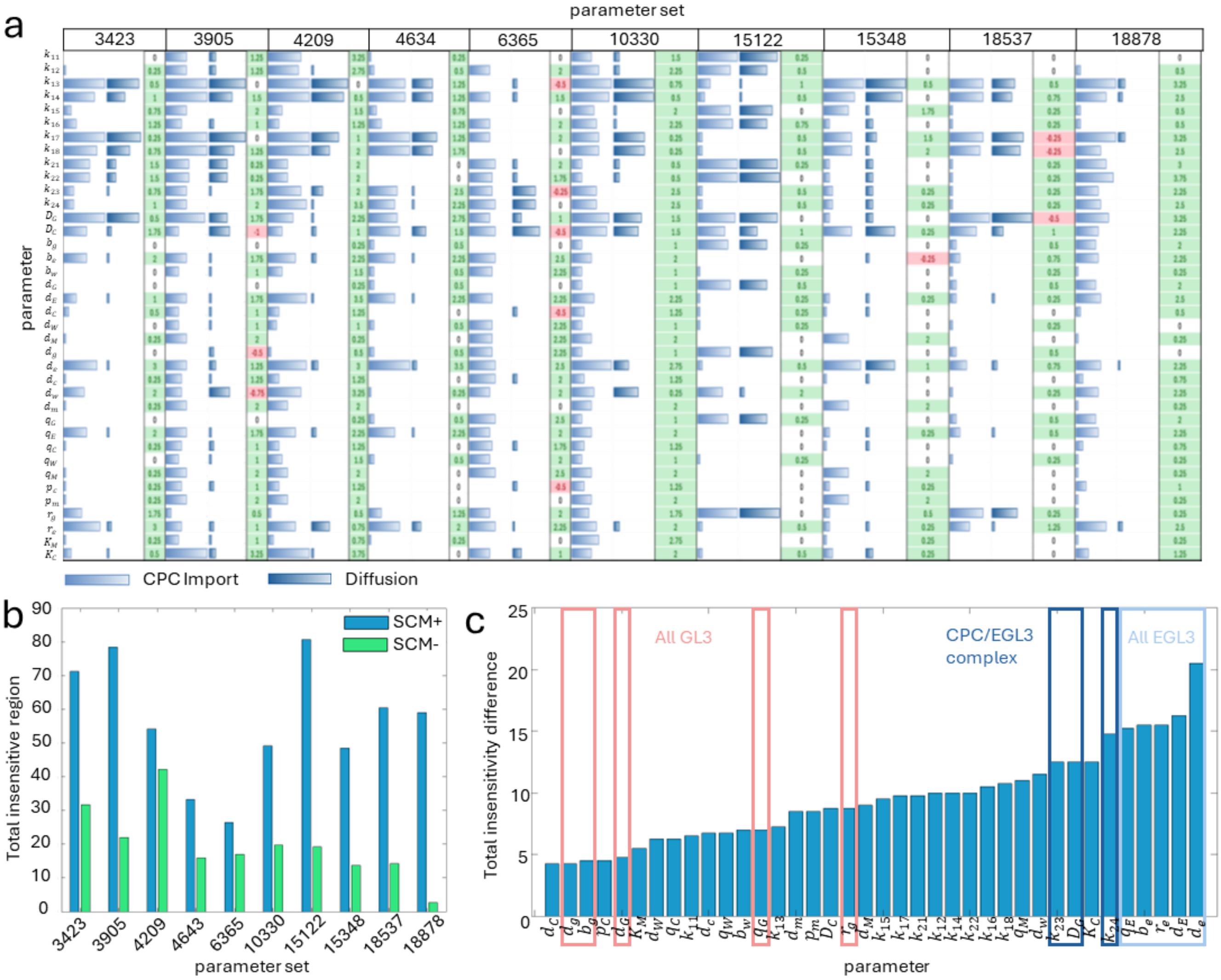

**Figure 4: Sensitivity analysis data. (a)** Insensitive regions for each core parameter, in each of the ten successful parameter sets found for the model with *WER* transcription repressed by cortical signal and the CPC complex. For each parameter set; the first column shows the insensitive regions found when using the SCM+ model (CPC import and diffusion), the second column shows insensitive regions found when using the SCM- model (diffusion), and the third column shows the difference between the insensitivity intervals of SCM+ and SCM- models, in orders of magnitude. Green indicates that the SCM+ model was less sensitive than the SCM- model to changes in the parameter, $p_{i,j}$, red indicated that the SCM+ model was more sensitive. **(b)** Total insensitive regions (units, orders of magnitude) for each of the ten successful parameter sets. **(c)** Total insensitivity difference for each parameter, summed over the ten successful parameter sets, and arranged in ascending order.

## Multiple binding site reactions within the MYB23 positive feedback loop are essential for successful root hair patterning.

Multiple binding sites have been reported at three locations within the *Arabidopsis* root hair transcription factor network. The *MYB23*[5], and *CPC* promotors[12,13], each contain multiple *WER* binding sites, consistent with cooperative transcriptional regulation, whilst WER/MYB23 complex formation is thought to be multimeric[9], implicating oligomerisation in the network. Models were solved with one, two, and all three of the multiple binding site reactions removed, (Table 7), to investigate their contribution to patterning.

The removal of cooperativity at the *MYB23* promotor was the only single multiple binding site removal which resulted in fewer successful parameter sets found relative to the full model. Furthermore, the *MYB23* promotor was the only multiple binding site reaction which, when alone, resulted in more successful parameter sets than the full model Fig. 5a (Supplementary File 5). These results suggest that *MYB23* promotor cooperativity is a key driver of *Arabidopsis* root hair patterning.

No successful parameter sets were found when all multiple binding site reactions were removed, nor when only *CPC* promotor cooperativity remained. Successful parameter sets were found when multiple binding sites were enabled on the *MYB23* promotor or in WER/MYB23 complex formation, either alone or in combination. As *MYB23* transcription occurs within a positive feedback loop with the WER/MYB23 complex, these results suggest that cooperativity or oligomerisation within this loop is essential to enable successful patterning.

Mathematically, increasing cooperativity or oligomerisation results in more pronounced nonlinearities in the corresponding reactions, Figs. 5b and 5c. Nonlinearities render a reaction insensitive at low regulator concentrations whilst enabling ultrasensitive, switch-like behaviour, as regulator concentrations increase. These dynamics would provide the MYB23 positive feedback loop with robustness against noise-driven patterning early in development[48,49] whilst promoting stable fate commitment at later stages.

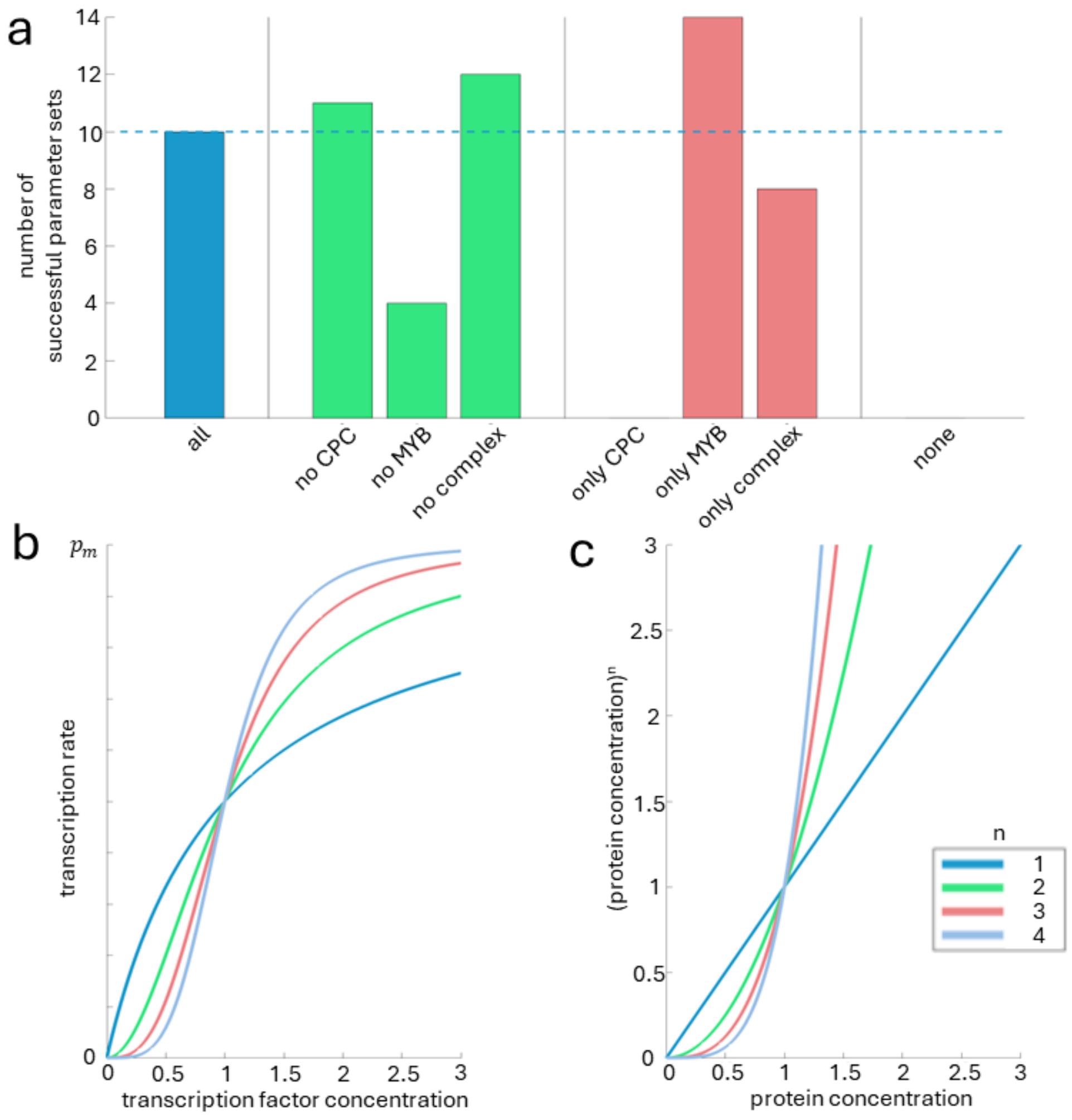

**Figure 5: Cooperativity and oligomerisation. (a)** The number of successful parameter sets found for the model with *WER* transcription repressed by cortical signal and the CPC complex, for each combination of multiple binding site reactions (Table 7). The blue bar shows the data when all three multiple binding sites are present, green bars show the data when only one multiple binding site reaction is removed, and red bars show the data when only one multiple binding site reaction remains. The dotted line is for comparison between the data for all three multiple binding sites and fewer than three multiple binding sites. **(b)** Relationship between transcription rate and transcription factor concentration when transcription cooperativity is present within a promotor. Formula, $p_m[TF]^n(1+[TF]^n)^{-1}$, where $p_m$ is the maximum transcription rate, $[TF]$ is transcription factor concentration, and $n=\{1,2,3,4\}$ is cooperativity. **(c)** A graph showing the relationship between the rate of protein complex formation and the concentration of protein. $n=\{1,2,3,4\}$ is number of proteins in the complex.

## Restricted EGL3 movement and CPC-EGL3, WER-GL3 preferential binding enable epidermal patterning.

EGL3 movement has been reported to be restricted relative to GL3[24]. In the models presented thus far, EGL3 was unable to move between epidermal cells, $D_E=0$. To understand whether restricted EGL3 movement is necessary for successful patterning, the models were solved with $D_E>0$. Two successful parameter sets were found, one unique and one previously identified when $D_E=0$. In both cases, the diffusion coefficient of GL3 exceeded that of EGL3 (Supplementary File 6). Moreover, when imposing equal mobility, $D_E=D_G$, all eleven successful parameter sets failed to reproduce experimental data. Next, the eleven parameter sets were solved with $D_E=10^{-\alpha}D_G$, for $\alpha=\{0.05,0.1,0.15,\ldots,3\}$ to find the maximum value of $D_E$ that enables patterning, $D_{E_{MAX}}$, (Supplementary File 6). In all cases $D_{E_{MAX}}<D_G$. Together, the data suggest that EGL3 movement must be restricted relative to GL3 for successful *Arabidopsis* root epidermal patterning.

The ability of GL3 to diffuse more quickly than EGL3 from trichoblast to atrichoblast would be of limited use if GL3 was sequestered before it could diffuse. Thus, it would be optimal for patterning if CPC bound EGL3 more strongly than GL3, and WER/MYB23 bound GL3 more strongly than EGL3. Binding affinity ratios supported CPC-EGL3 ($\approx 73\%$) and WER-GL3 ($\approx 64\%$) preferential binding in the majority of successful parameter sets, (Supplementary File 6). Only a minority ($\approx 36\%$) satisfied MYB23 binding GL3 more strongly than EGL3, suggesting that preferential binding of MYB23-GL3 is less important than WER-GL3. This distinction is likely due to the WER complex concentration being greater than the MYB23 complex concentration in atrichoblasts during early development, as the *MYB23* promotor contains multiple WER binding sites[5] and the WER complex must become established before MYB23 is transcribed, Fig. 5b. Later, once the MYB23 positive feedback loop dominates in atrichoblast cells, MYB23 abundance alone is enough to sequester the majority of GL3, leaving CPC to be imported into trichoblast cells. Further analysis showed that the number of optimal relationships satisfied by a parameter set correlated positively (r = $0.67$, p = $0.02$) with $-\alpha$, the order of magnitude difference between the EGL3 and GL3 diffusion coefficients, Fig. 6a. Thus, the restriction of EGL3 movement needed for correct patterning is dependent on other parameters, if parameters are set such that they aid complex accumulation in the correct cell type then EGL3 can move more freely.

Consistent with these relationships, pooling all $220$ SCM+ steady state solutions revealed that the CPC/EGL3 complex was more abundant than CPC/GL3 in trichoblast cells and WER/GL3, MYB23/GL3 were more abundant than WER/EGL3, MYB23/EGL3 in atrichoblast cells, Fig. 6b. The dominant complex in trichoblast cells was CPC/EGL3, aligning with sensitivity analysis results, which identified CPC-EGL3 related parameters as most affected by removal of SCM-mediated CPC import. In atrichoblasts, whilst there was a large abundance of MYB23/GL3 complex as expected, there was less WER complex than CPC complex. This result shows that, while WER is important in initiating correct epidermal patterning, once the MYB23 positive feedback loop is activated WER becomes redundant.

Collectively, these data show correct epidermal patterning relies on the combined effects of restricted EGL3 movement and biased complex formation, enabling spatial segregation of cell fates despite shared network components.

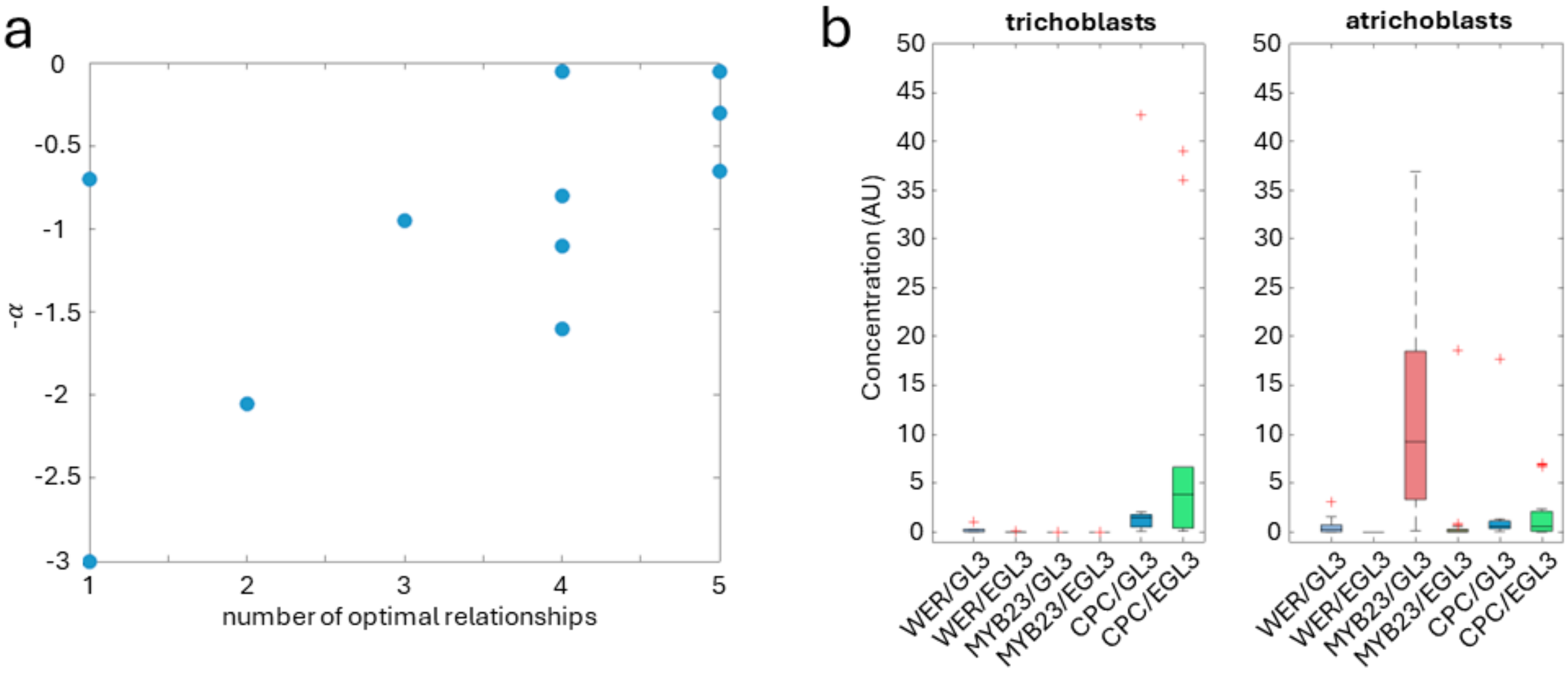

**Figure 6: EGL3 movement. (a)** Boxplots showing concentrations of protein complexes in trichoblast and atrichoblast cells, for all 220 SCM+ steady state solutions, with $D_E = D_{E_{MAX}}$, for the SCM+ model with *WER* translation regulated by the CPC complex and cortical signal. **(b)** Scatterplot showing the correlation between the number of optimal relationships a parameter set satisfied, Table 10, and the relationship between $D_G$ and $D_{E_{MAX}}$. $D_{E_{MAX}} = 10^{-\alpha} D_G$, Table 9. One data point per parameter set.

## Trichoblast patterning is controlled by combined activator inhibitor and substrate depletion mechanisms.

To gain a holistic understanding of the reaction network patterning *Arabidopsis* root hairs, the network was studied in the context of fundamental patterning mechanisms[50,51].

The SCM- model is not a classical reaction-diffusion model because cortical signal influences *WER* transcription in H positioned epidermal cells [3], Equation (5). Positional bias was removed from the SCM- model, $(X_i = 0 \; \forall \; i)$, to test whether the reaction network contains a spontaneously patterning reaction-diffusion system.

As expected, the SCM-, no cortical signal, reaction-diffusion model did not reproduce *scm* mutant data, which is biased towards trichoblast cells in H positions. However, the SCM- no cortical signal model could pattern into distinct trichoblast and atrichoblast cell types, as defined by the colocalization of network components characteristically expressed in each cell type. This result shows that the core reaction network patterning *Arabidopsis* root hairs is a spontaneously patterning reaction-diffusion system.

Two network interactions, newly discovered here, namely a nonlinearity in the MYB23 positive feedback loop and *WER* transcriptional repression by a network component colocalised with the CPC complex, form an activator-inhibitor patterning motif[50,51], where the activator is the MYB23 complex and the inhibitor is the CPC complex, Fig. 7, green and red connectors. Conditions necessary for activator-inhibitor systems to pattern are, (i) a nonlinearity in activator positive feedback and, (ii) the inhibitor must diffuse faster than the activator. Both of these conditions have been shown to be satisfied in the *Arabidopsis* root hair patterning network, the nonlinearity was shown to be necessary for correct patterning, and CPC, an essential component of the inhibitor complex, can diffuse between cells[5,18] whereas MYB23, an essential component of the activator complex, does not [5].

Activator-inhibitor dynamics alone do not fully describe the behaviour of the transcription factor network. GL3 and EGL3 are essential for all regulatory interactions in the spontaneously patterning reaction-diffusion system, Fig. 7, and these are transcriptionally repressed via the activator, indicating that a second classical patterning mechanism, substrate depletion[51], also operates within the network. In the *Arabidopsis* root epidermis, the nonlinear MYB23 positive feedback loop depends on MYB23 complex formation and thus on GL3/EGL3, the substrate. In turn, MYB23 complex inhibits the local *GL3/EGL3* transcription[5], limiting substrate availability. Removal of substrate depletion, i.e. WER complex inhibition of either *EGL3* or *GL3* transcription, from the SCM- no cortical signal model prevented *in-silico* patterning, showing that substrate depletion is essential to *Arabidopsis* root epidermal patterning, Fig. 7, green and yellow connectors.

These results show that correct *Arabidopsis* trichoblast patterning relies upon the coordinated action of both an activator-inhibitor and a substrate depletion mechanism which overlap in their requirement for nonlinear promotion of the activator complex.

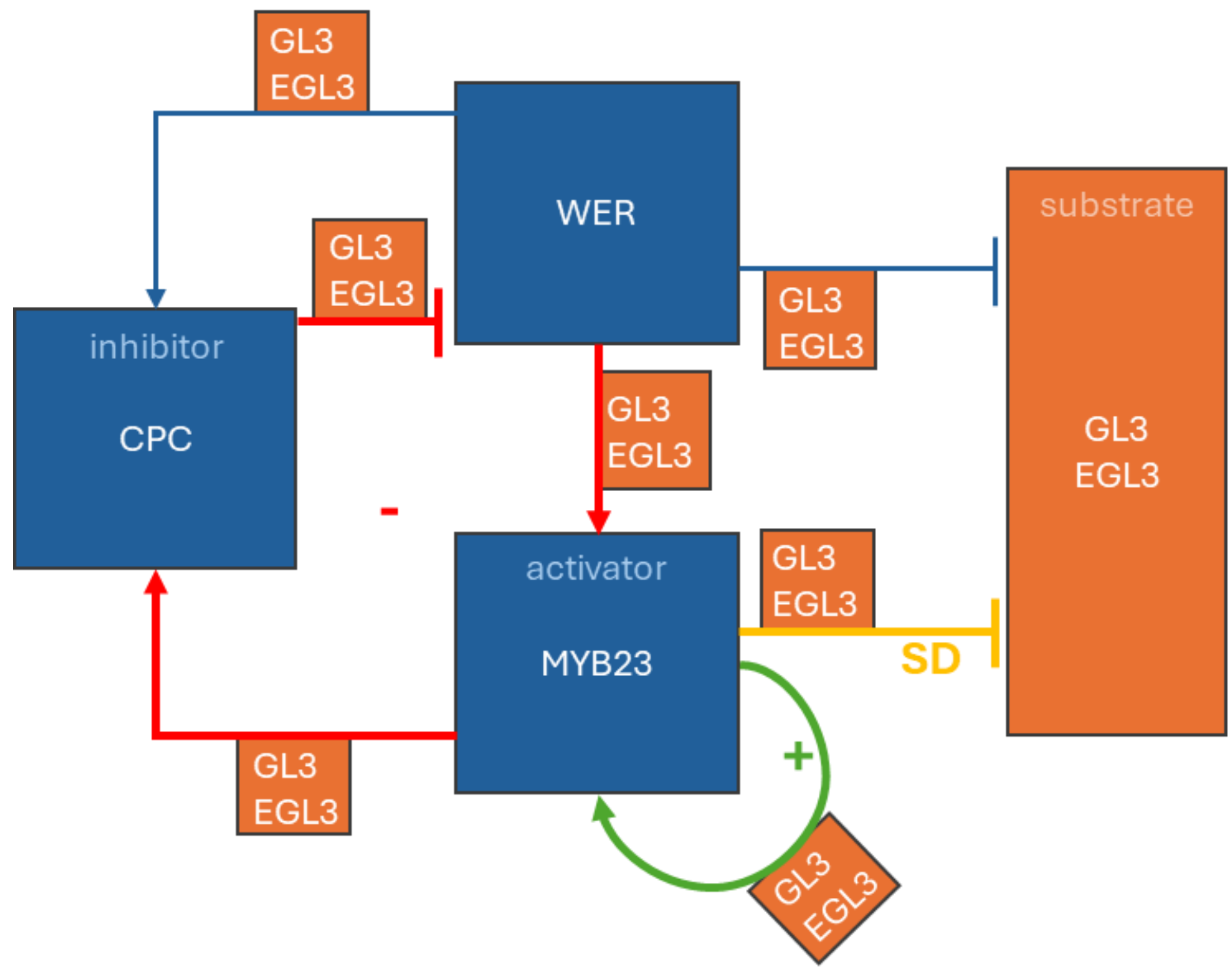

**Figure 7: *Arabidopsis* root epidermis reaction-diffusion network diagram.** Arrows represent transcriptional promotion or protein binding. Barbed arrows represent transcriptional repression. Red connectors indicate that the reactions are part of the activator-inhibitor patterning mechanism. Yellow connectors indicate that the reactions are part of the substrate depletion patterning mechanism. Green connectors indicate the nonlinear positive feedback on the activator which is key to both the activator-inhibitor and substrate depletion patterning mechanisms. CPC and GL3 diffuse. All reactions depend on complex formation, illustrated by GL3/EGL3 on all connectors.

# Discussion

Many components of the *Arabidopsis* root epidermal patterning transcription factor network are well characterised. However, the network's complexity has made it difficult to understand how the regulatory architecture generates robust epidermal patterning. The work here aimed to bridge that gap. Mathematical modelling was used to integrate the wealth of experimental data into a single transcription factor network model. By maintaining a one-to-one relationship between the model and the biological network, the sufficiency of current knowledge to explain wild-type and mutant patterning could be systematically tested, and essential regulatory features identified.

## Results summary.

Modelling results suggested that the current understanding of *WER* transcriptional regulation was incomplete. Previously hypothesised mechanisms[10,9] were insufficient to reproduce epidermal patterning, indicating that additional regulation of *WER* transcription is required. WER autoregulation had previously been excluded experimentally[10,5], thus, it was hypothesised that successful patterning may be achieved if *WER* transcription was repressed by a transcription factor network component dominant in trichoblast cells. Introducing *WER* transcriptional repression by the CPC complex to the model enabled *in-silico* reproduction of wild-type and mutant patterns across real *Arabidopsis* topologies. Subsequent mechanistic investigations were performed with all three *WER* regulation mechanisms, and all but one investigation required *WER* regulation by the CPC complex to reproduce experimental data. The exception arose when unbound CPC repressed *WER* and MYB23 cooperativity was the only multiple binding site retained. This model yielded far fewer successful parameter sets than *WER* repression by the CPC-complex and is less plausible given that CPC is unlikely to bind the WER promotor directly. Indeed, promotor sequence analysis supported the viability of CPC-complex *WER* regulation.

In recent years the role of SCM has been revised from a cortical signal receptor[20] to a CPC importer[17]. While removal of a cortical signal receptor would be expected to decouple epidermal patterning from positional cues, it was not obvious that loss of CPC import, while leaving positional cues intact, could produce a positionally impaired pattern. Here, it was shown that the removal of CPC import was sufficient to reproduce *scm* mutant data. Furthermore, SCM-mediated CPC import increased patterning robustness to reaction rate variation by ensuring robust formation of the CPC/EGL3 complex.

The functional contribution of multiple binding site reactions within the network was examined. No benefit to successful patterning was found for cooperativity on CPC transcriptional regulation. In contrast, a multiple binding site reaction within the MYB23 positive feedback loop was essential, with cooperativity on MYB23 transcription contributing more strongly to successful patterning than WER/MYB23 complex oligomerisation. These results suggest that amplification of the central transcription factor complex must be protected against noise driven patterning early in development whilst exhibiting switch-like behaviour to promote distinct cell fates when regulator concentrations increase.

In addition to differential binding preferences, asymmetric mobility of GL3 and EGL3 was essential for patterning. Previous work showed that GL3 moves into atrichoblast cells where it accumulates[11], whereas EGL3 remains largely restricted to trichoblast cells[24]. Patterning could be reproduced *in-silico* when EGL3 diffused between cells, provided its diffusion rate remained lower than that of GL3. Favourable steady-state relationships for correct patterning were defined as; CPC preferentially binding EGL3, WER and MYB23 preferentially binding GL3; CPC/EGL3 dominating in trichoblast cells; and MYB23/GL3 dominating in atrichoblast cells. Satisfaction of these relationships reduced the required difference between GL3 and EGL3 mobility while still allowing correct patterning.

Combining the results of this study with published data uncovered a reaction-diffusion patterning network at the core of *Arabidopsis* root epidermal patterning. Indeed, epidermal cells spontaneously patterned *in-silico* in the absence of positional signalling and SCM. The reaction-diffusion network was shown to contain two classic patterning mechanisms, activator-inhibitor and substrate depletion, both of which were essential to achieve distinct epidermal cell fates.

## The activator-inhibitor and substrate depletion patterning mechanisms work together to pattern the *Arabidopsis* root epidermis.

As epidermal cells emerge from the root apical meristem, they produce WER and GL3/EGL3, enabling formation of the WER complex. Biological noise generates cell-cell differences in WER complex concentration. As the WER complex promotes *CPC* and represses *GL3/EGL3* expression, CPC and GL3 concentration gradients emerge between neighbouring cells, Figs. 8a and 8b. These asymmetries are amplified by the substrate depletion mechanism. WER complex promotion of *MYB23* causes MYB23 and CPC concentrations to peak in the same cell, where they compete to bind the substrate GL3/EGL3. Once MYB23 complex levels exceed some threshold, the nonlinear positive feedback loop allows MYB23 to increasingly represses GL3/EGL3 expression, limiting available substrate and outcompete CPC for the limited resource. Unbound CPC is thus available to diffuse into neighbouring cells, where it promotes trichoblast fate, Fig 8c. As MYB23 can substitute for WER in downstream regulation, WER becomes redundant after establishing the initial CPC and GL3 gradients.

The activator-inhibitor mechanism operates in parallel with the substrate depletion mechanism to prevent MYB23 positive feedback from becoming established in all cells, Figs. 8c and 8d. Restricted EGL3 mobility works to maintain a local pool of substrate in trichoblast cells, enabling CPC complex formation. The CPC complex represses *WER* transcription and prevents activation of the MYB23 positive feedback loop Figs. 8d to 8f.

At its core, the *Arabidopsis* root hair transcription factor network is a reaction-diffusion patterning network containing two classical patterning motifs, activator-inhibitor and substrate depletion. In the root epidermis the reaction-diffusion network is biased by a currently unidentified positional signal. Positional signalling overrides biological noise and reduces *WER* expression in H positioned cells, dictating trichoblast positioning over cortical clefts. In parallel, SCM accumulates in H positioned cells importing CPC from neighbouring N positioned cells, reinforcing CPC flux in the same direction as diffusion. The removal of CPC from N positioned cells reduces CDC competition with WER/MYB23 for GL3/EGL3, reinforcing cell fait determination. Consistent with this, sensitivity analysis showed that directed CPC import reduces network sensitivity to reaction rate changes.

Together, these results reveal that the *Arabidopsis* root hair transcription factor network is a spontaneously patterning reaction-diffusion system containing integrated activator-inhibitor and substrate depletion mechanisms. The work presented here provides an intuitive understanding of how intracellular feedback, intercellular transport, and positional cues combine to ensure robust root hair patterning.

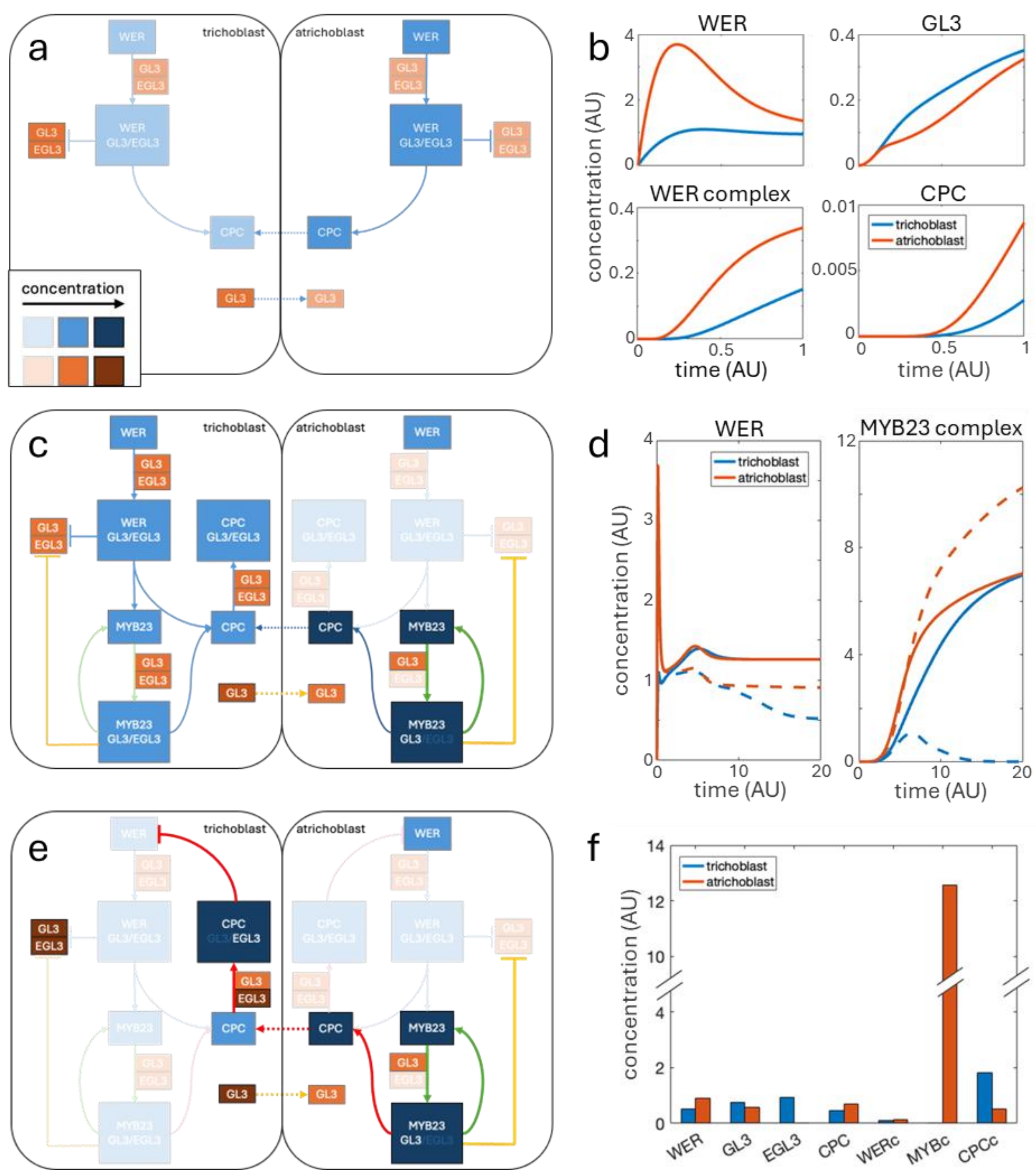

**Figure 8: Reaction-diffusion network patterning the *Arabidopsis* root epidermis.** In panels a c and e Connector representations are the same as in Figure 7. Dotted arrows show the direction of diffusive flux. The darker the colour of the connecter the more impact the reaction has on patterning. The darker the colour of network components that higher their concentration. In panels b d and f blue shows concentrations in trichoblast cells, red shows concentrations in atrichoblast cells. **(a)** Initial differences in WER complex concentration between neighbouring cells produces CPC and GL3 concentration gradients. **(b)** Model solutions early in development, showing concentrations of components in panel A. **(c)** The interaction network without the activator-inhibitor mechanism, there is nothing to stop the positive feedback loop from becoming established in the trichoblast cell. **(d)** Model solutions showing that initial differences in WER and MYB23 concentrations are lost in the reaction network without the activator-inhibitor mechanism, solid lines. Dotted lines show model solutions when the activator-inhibitor mechanism is present. **(e)** Fully differentiated interaction network. **(f)** Steady state concentration of proteins and protein complexes.

# Research Funding.

Funded by the Leverhulme Project Grant RPG-2021-053.

# References.

# File S1

Supporting information for Methods Sections, 'Model Description' and 'Parameter Search' and Results Section, 'The transcription factor network model can reproduce wild-type and *scm* mutant epidermal patterning if repression of *WER* is colocalised with the CPC complex' of the main manuscript. WT images available on GitHub[1].

**Table S1.** Mappings of cortical clefts and epidermal cells for twenty-two root cross-sections. Nineteen distinct cortical cleft, epidermal cell relationships were found. 'H' and 'N', represent an epidermal cell in the H or N position, i.e. 'H' shows an epidermal cell overlying a cortical cleft. Blue lines under letters show GL2 expression within the epidermal cell. The notation '1(=16)', for example, shows topologically equivalent root cross-sections.

| root # | | | | | | | | | | | | | | | | | | | | | | | # of epidermal cells |
|---|---|---|---|---|---|---|---|---|---|---|---|---|---|---|---|---|---|---|---|---|---|---|---|
| 19 | H | N | H | N | H | N | H | N | N | H | N | H | N | H | N | N | H | N | | | | | 18 |
| 1(=16) | H | N | H | N | H | N | N | H | N | H | N | N | H | N | N | H | N | H | N | | | | 19 |
| 3 | H | N | H | N | H | N | N | H | N | H | N | H | N | H | N | N | H | N | N | | | | 19 |
| 5 | H | H | N | N | H | N | H | N | H | N | H | N | N | H | N | N | H | N | N | | | | 19 |
| 6 | H | N | H | N | H | N | H | N | H | N | H | N | N | H | N | N | H | N | N | | | | 19 |
| 8 | H | N | H | N | N | H | N | N | H | N | H | N | N | H | N | H | H | N | N | | | | 19 |
| 16(=1) | H | N | N | H | N | H | N | H | N | H | N | H | N | N | H | N | H | N | N | | | | 19 |
| 20 | H | N | N | H | N | H | N | H | N | N | H | N | H | N | H | N | N | H | N | | | | 19 |
| 2 | H | N | N | H | N | N | H | N | H | N | N | H | N | N | H | N | H | N | H | N | | | 20 |
| 9(=14) | H | N | N | H | N | H | N | H | N | H | N | N | H | N | N | H | N | N | H | N | | | 20 |
| 10(=11) | H | N | N | H | N | N | H | N | H | N | H | N | N | H | N | H | N | N | H | N | | | 20 |
| 11(=10) | H | N | N | H | N | N | H | N | H | N | H | N | N | H | N | H | N | N | H | N | | | 20 |
| 12 | H | N | N | H | N | H | N | N | H | N | H | N | N | H | N | H | N | H | N | N | | | 20 |
| 13 | H | N | N | H | N | H | N | H | N | N | H | N | H | N | H | N | N | H | N | N | | | 20 |
| 14(=9) | H | N | H | N | H | N | N | H | N | N | H | N | N | H | N | H | N | N | H | N | | | 20 |
| 17 | H | N | H | N | H | N | N | H | N | H | N | N | H | N | N | H | N | H | N | N | | | 20 |
| 4 | H | N | N | H | N | N | H | N | H | N | H | N | H | N | N | H | N | N | H | N | N | | 21 |
| 7 | H | N | N | H | N | N | N | H | N | H | N | H | N | N | H | N | N | N | H | N | N | | 21 |
| 21 | H | N | N | H | N | N | H | N | N | H | N | N | H | N | H | N | H | N | N | H | N | | 21 |
| 22 | H | N | N | H | N | N | H | N | H | N | H | N | N | H | N | N | H | N | H | N | N | | 21 |
| 15 | H | N | N | H | N | H | N | H | N | N | H | N | N | H | N | N | H | N | N | H | N | N | 22 |
| 18 | H | N | H | H | N | H | N | N | H | N | N | H | N | H | N | N | N | H | N | H | N | N | 22 |

| representative | H | N | H | N | H | N | H | N | N | H | N | N | H | H | N | N | H | N | N | 19 cells |
|---|---|---|---|---|---|---|---|---|---|---|---|---|---|---|---|---|---|---|---|---|

**S1 Refences.**

# File S2

Supporting information for Methods Section, 'Parameter Search' of main manuscript.

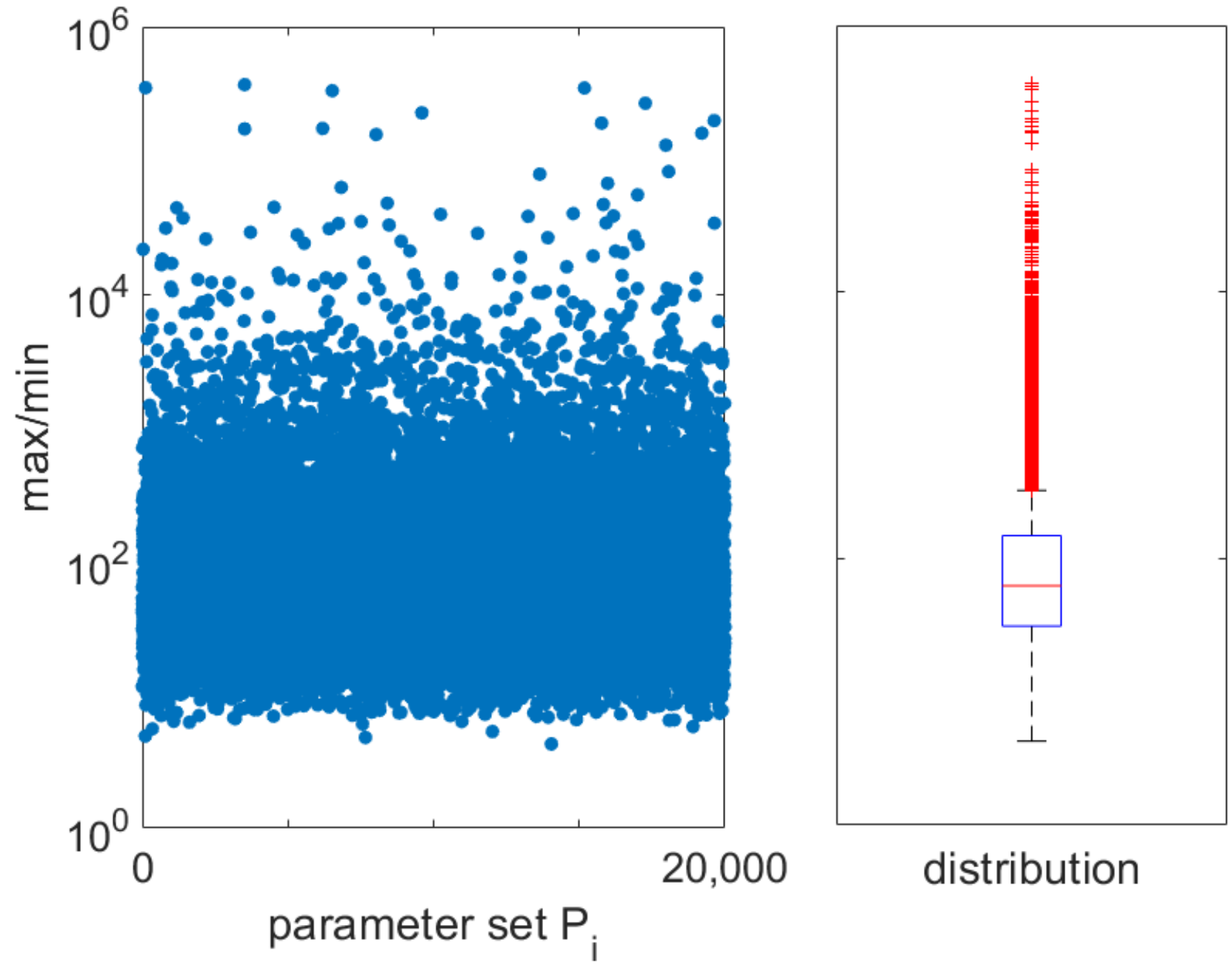

**Figure S2-1. The maximum order of magnitude difference for each parameter set.** Twenty thousand sets of forty-three parameter values are chosen uniformly at random from the interval $(0,10]$. **Left.** The maximum of the forty-three parameter values divided by the minimum of the forty-three parameter values, for each of the twenty thousand parameter sets. **Right.** A box plot of the individual parameter set data.

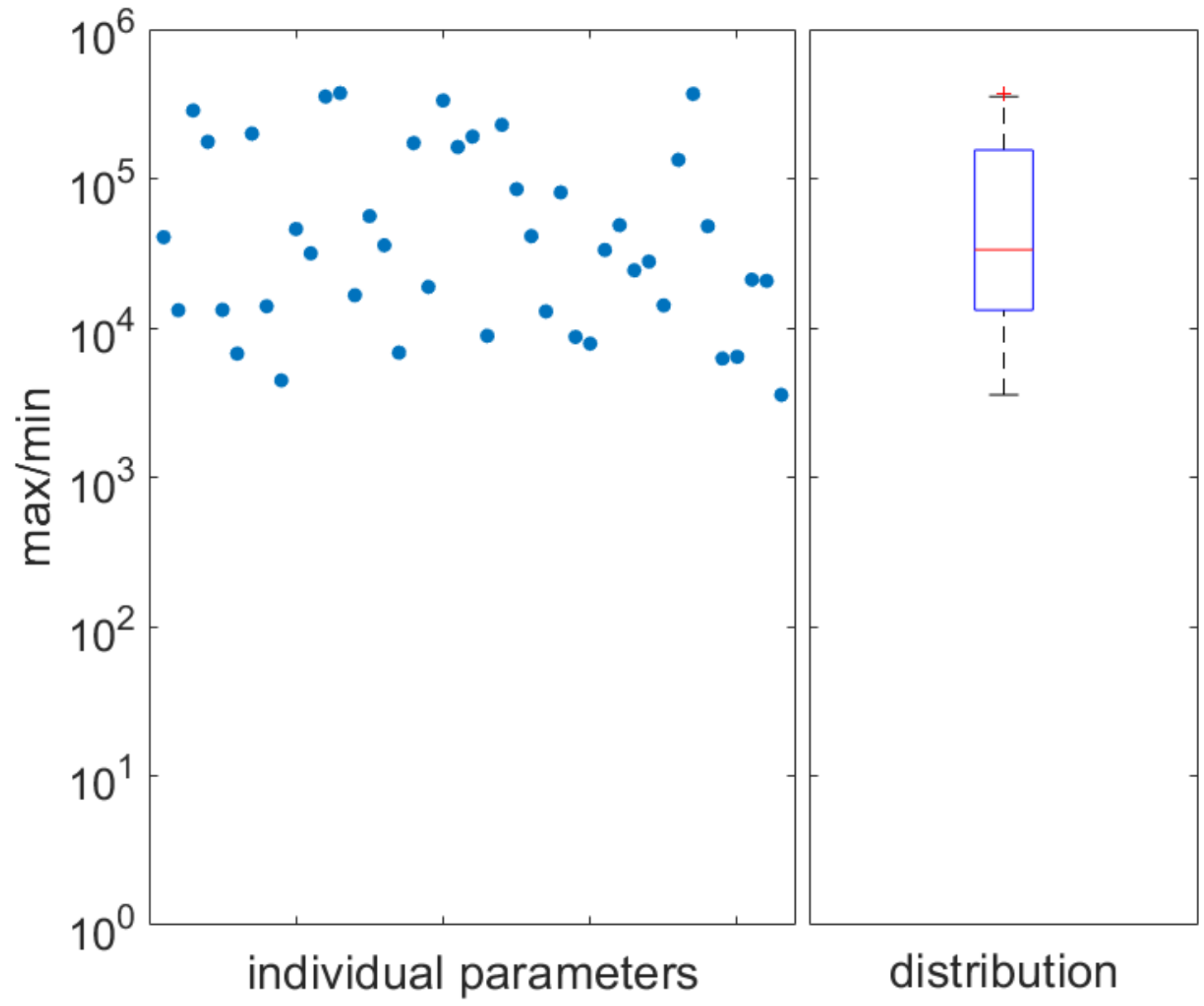

**Figure S2-2. The maximum order of magnitude difference for each parameter.** Forty-three parameter values are chosen uniformly at random from the interval $(0,10]$. Twenty thousand choices were made for each parameter. **Left.** The maximum of the twenty thousand values for a parameter divided by the minimum of the twenty thousand values, for each of the forty-three parameters. **Right.** A box plot of the individual parameter data.

**Table S2-1.** Analysis of wildtype data calculated from GL2mT images (S1 File). WT and *scm* mutant image analysis available on GitHub [1].

| root # | atrichoblast | | trichoblast | |
|---|---|---|---|---|
| | H position | N position | H position | N position |
| 1 | 87.5 | 12.5 | 0.0 | 100.0 |
| 2 | 100.0 | 0.0 | 0.0 | 100.0 |
| 3 | 100.0 | 0.0 | 0.0 | 100.0 |
| 4 | 100.0 | 0.0 | 0.0 | 100.0 |
| 5 | 100.0 | 0.0 | 0.0 | 100.0 |
| 6 | 87.5 | 12.5 | 0.0 | 100.0 |
| 7 | 85.7 | 14.3 | 0.0 | 100.0 |
| 8 | 75.0 | 25.0 | 9.1 | 90.9 |
| 9 | 87.5 | 12.5 | 8.3 | 91.7 |
| 10 | 100.0 | 0.0 | 0.0 | 100.0 |
| 11 | 100.0 | 0.0 | 8.3 | 91.7 |
| 12 | 100.0 | 0.0 | 0.0 | 100.0 |
| 13 | 100.0 | 0.0 | 0.0 | 100.0 |
| 14 | 100.0 | 0.0 | 0.0 | 100.0 |
| 15 | 87.5 | 12.5 | 0.0 | 100.0 |
| 16 | 75.0 | 25.0 | 0.0 | 100.0 |
| 17 | 100.0 | 0.0 | 0.0 | 100.0 |
| 18 | 88.9 | 11.1 | 0.0 | 100.0 |
| 19 | 100.0 | 0.0 | 0.0 | 100.0 |
| 20 | 100.0 | 0.0 | 0.0 | 100.0 |
| 21 | 100.0 | 0.0 | 0.0 | 100.0 |
| 22 | 87.5 | 12.5 | 0.0 | 100.0 |
| **mean** | 93.7 | 6.3 | 1.2 | 98.8 |
| **standard deviation** | 8.4 | 8.4 | 3.0 | 3.0 |

**Table S2-2.** *scm* mutant data calculated from eleven figures within six publications [2–7].

| reference | figure | mutant | marker | trichoblast | | atrichoblast | |
|---|---|---|---|---|---|---|---|
| | | | | **H position** | **N position** | **H position** | **N position** |
| Kwak 2005 [2] | 1A | scm-1 | GL2 | 66.7 | 33.3 | 23.6 | 76.4 |
| Kwak 2005 [2] | 1E | scm-2 | GL2 | 54.5 | 45.5 | 36.0 | 64.0 |
| Kwak 2005 [2] | 2A | scm-2 | CPC | 57.1 | 42.9 | 15.0 | 85.0 |
| Kwak 2005 [2] | 2B | scm-2 | WER | 85.7 | 14.3 | 7.7 | 92.3 |
| Kwak 2005 [2] | 2D | scm-2 | EGL3 | 72.5 | 27.5 | 11.3 | 88.7 |
| Kwak 2007 [3] | 2A | scm-2 | GL2 | 47.6 | 52.4 | 13.0 | 87.0 |
| Kwak 2007 [3] | 7B | scm-2 | GL2 | 39.3 | 60.7 | 40.0 | 60.0 |
| Song 2011 [4] | 2B | scm-2 | GL2 | 66.7 | 33.3 | 6.3 | 93.7 |
| Kwak 2014 [5] | 2B | scm-2 | GL2 | 55.6 | 44.4 | 37.0 | 63.0 |
| Song 2019 [6] | 2b | scm-2 | GL2 | 72.4 | 27.6 | 48.3 | 51.7 |
| Chaudhary 2021 [7] | 3J | sub-9 | GL2 | 61.5 | 38.5 | 8.7 | 91.3 |
| | | | **mean** | 61.8 | 38.2 | 22.4 | 77.6 |
| | | | **standard deviation** | 12.9 | 12.9 | 15.2 | 15.2 |

**S2 Refences.**

# File S3

Supporting information for 'Experimental Methods' Section of main manuscript.

**Table S3.** Primers used for cloning fragments for the GLABRA2 genomic fragment and Gly-Ser linker-mTurquoise fusion protein. Primers are expressed as 5'-3' orientation.

| Primer name | Primer sequence |
|---|---|
| GL2 B Fwd 1 | aacaggtctcaAACAATGAAGTCGATCGATGGCTG |
| GL2 B Fwd 2 | aacaggtctcaGGTATCCGTGGAGGACAG |
| GL2 B Rev 1 | aacaggtctcaTACCGAGACGTCCACTATTG |
| GL2 B Rev 2 | aacaggtctcaAGCCGCAATCTTCGATTTGTAGAC |
| Linker-mTurq C Fwd | aacaggtctcaGGCTtccggaggtggtggttctggtGGGGGTGGCTCTgtgagcaagggcgaggag |
| mTurq C Rev | aacaggtctcaCTGAttactcttcttcttgatcagcttctgtg |

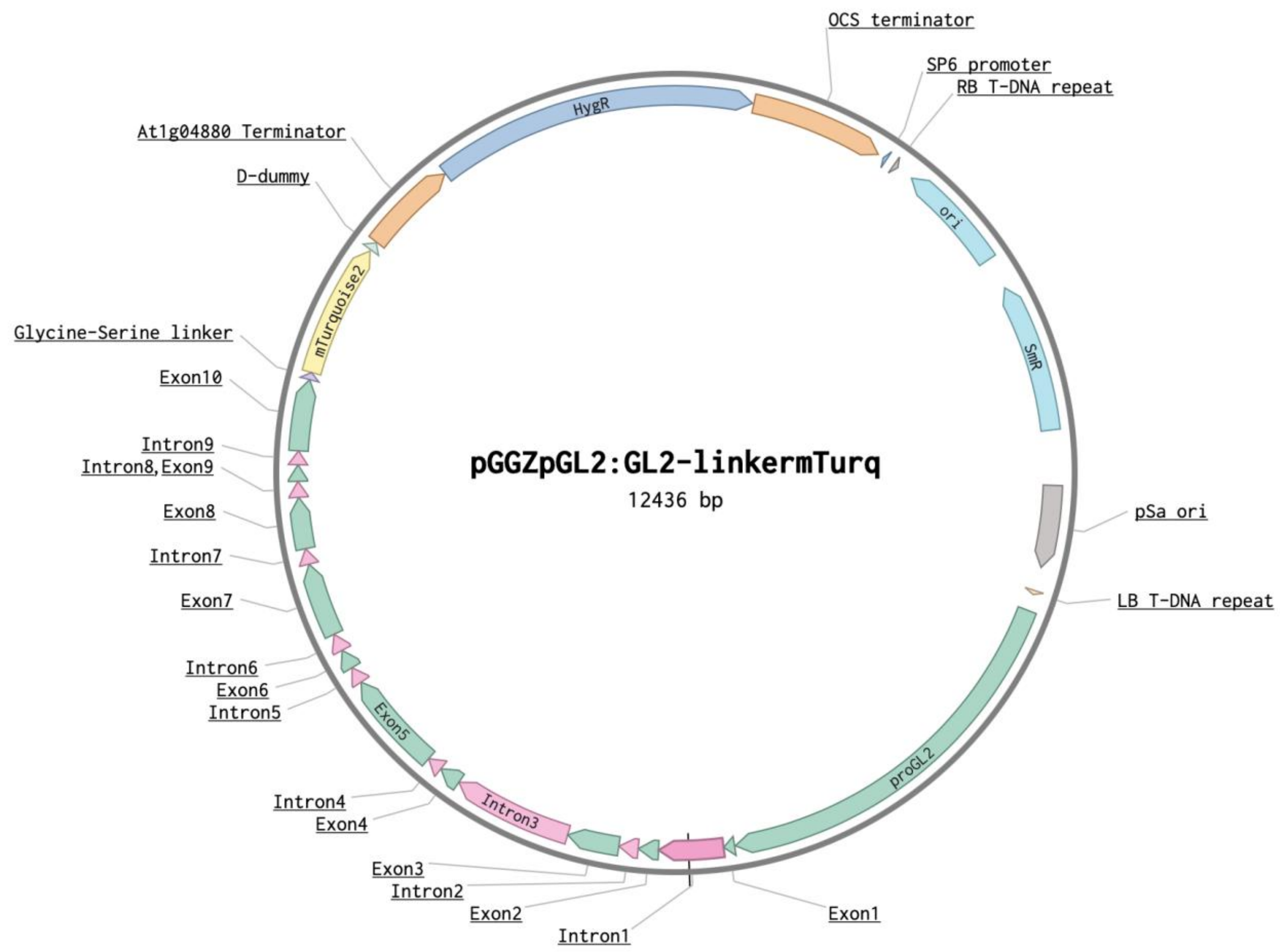

**Figure S3.** Plasmid map of GL2-mTurquoise reporter construct as sequenced. Total size of plasmid 12436 base pairs.

Acknowledgement. The Salk Institute Genomic Analysis Laboratory provided the sequence-indexed Arabidopsis T-DNA insertion mutants.

# File S4

Supporting information for Results Section, 'The transcription factor network model can reproduce wild-type and *scm* mutant epidermal patterning if repression of *WER* is colocalised with the CPC complex' of main manuscript.

**Table S4-1.** Parameter values for the 10 successful parameter sets, given to two decimal places.

| | **parameter set number** | **3423** | **3905** | **4209** | **4634** | **6365** | **10330** | **15122** | **15348** | **18537** | **18878** |
|---|---|---|---|---|---|---|---|---|---|---|---|
| $\boldsymbol{k_{11}}$ | $A_{GW}$ association | 5.95 | 4.92 | 4.61 | 0.82 | 0.28 | 7.36 | 3.38 | 8.57 | 2.07 | 5.66 |
| $\boldsymbol{k_{12}}$ | $A_{GW}$ disassociation | 1.67 | 4.80 | 0.46 | 6.79 | 4.14 | 9.09 | 2.82 | 8.85 | 5.96 | 5.47 |
| $\boldsymbol{k_{13}}$ | $A_{EW}$ association | 5.29 | 2.31 | 2.34 | 3.18 | 1.10 | 0.17 | 8.71 | 1.48 | 3.19 | 3.40 |
| $\boldsymbol{k_{14}}$ | $A_{EW}$ disassociation | 3.66 | 4.52 | 6.31 | 4.80 | 3.46 | 3.30 | 4.97 | 3.74 | 8.53 | 9.89 |
| $\boldsymbol{k_{15}}$ | $A_{GM}$ association | 1.50 | 2.28 | 0.42 | 4.03 | 0.93 | 2.46 | 7.57 | 0.26 | 2.59 | 7.55 |
| $\boldsymbol{k_{16}}$ | $A_{GM}$ disassociation | 0.09 | 6.54 | 0.81 | 1.74 | 2.63 | 5.71 | 3.54 | 5.02 | 9.59 | 1.63 |
| $\boldsymbol{k_{17}}$ | $A_{EM}$ association | 2.00 | 3.27 | 3.25 | 4.09 | 3.48 | 5.87 | 9.13 | 7.46 | 1.01 | 1.83 |
| $\boldsymbol{k_{18}}$ | $A_{EM}$ disassociation | 7.63 | 8.51 | 3.13 | 1.20 | 0.90 | 7.00 | 1.61 | 0.38 | 8.33 | 1.69 |
| $k_{21}$ | $I_{GC}$ association | 2.68 | 1.23 | 9.23 | 6.53 | 9.73 | 5.55 | 5.82 | 4.63 | 2.66 | 0.60 |
| $k_{22}$ | $I_{GC}$ disassociation | 6.21 | 7.82 | 8.67 | 2.07 | 6.32 | 4.23 | 7.03 | 3.94 | 7.27 | 1.98 |
| $k_{23}$ | $I_{EC}$ association | 9.98 | 9.62 | 6.51 | 5.66 | 6.28 | 6.94 | 9.18 | 7.77 | 6.64 | 7.58 |
| $k_{24}$ | $I_{EC}$ disassociation | 0.61 | 1.00 | 5.74 | 4.09 | 8.26 | 3.07 | 1.17 | 2.36 | 4.13 | 0.21 |
| $d_G$ | GL3 protein | 0.60 | 1.17 | 1.07 | 2.25 | 1.78 | 4.40 | 9.54 | 2.97 | 3.18 | 1.97 |
| $d_E$ | EGL3 protein | 2.38 | 3.34 | 6.72 | 5.98 | 6.99 | 9.55 | 1.82 | 6.70 | 3.18 | 6.12 |
| $d_C$ | CPC protein | 3.30 | 1.74 | 0.99 | 5.46 | 1.46 | 3.23 | 4.44 | 7.71 | 3.84 | 4.86 |
| $d_W$ | WER protein | 9.57 | 6.03 | 1.06 | 5.43 | 5.71 | 5.82 | 9.20 | 3.83 | 1.74 | 6.71 |
| $d_M$ | MYB23 protein | 6.54 | 8.03 | 1.76 | 0.84 | 4.67 | 1.16 | 1.93 | 9.98 | 2.85 | 4.78 |
| $d_g$ | *GL3* mRNA | 8.17 | 3.94 | 2.41 | 7.09 | 6 | 9.18 | 7.17 | 4.23 | 8.93 | 0.82 |
| $d_e$ | *EGL3* mRNA | 2.55 | 1.45 | 6.51 | 5.16 | 8.53 | 3.03 | 5.18 | 0.60 | 7.13 | 6.61 |
| $d_c$ | *CPC* mRNA | 7.45 | 6.77 | 3.32 | 7.81 | 8.76 | 3.33 | 4.71 | 6.73 | 3.89 | 3.99 |
| $d_w$ | *WER* mRNA | 8.47 | 5.97 | 3.25 | 1.20 | 3.29 | 0.43 | 0.95 | 4.51 | 4.77 | 5.91 |
| $d_m$ | *MYB23* mRNA | 1.68 | 6.07 | 7.58 | 4.15 | 1.50 | 1.77 | 8.87 | 0.40 | 6.83 | 3.63 |
| $b_g$ | *GL3* mRNA | 8.27 | 5.41 | 8.37 | 8.02 | 4.61 | 7.66 | 7.40 | 1.80 | 9.23 | 0.68 |
| $b_e$ | *EGL3* mRNA | 6.36 | 1.07 | 4.53 | 6.44 | 9.98 | 2.45 | 3.90 | 1.86 | 4.60 | 2.61 |
| $b_w$ | *WER* mRNA | 6.90 | 9.75 | 4.39 | 7.48 | 6.10 | 6.75 | 7.01 | 4.52 | 8.21 | 8.62 |
| $p_c$ | *CPC* mRNA | 5.64 | 6.01 | 8.22 | 4.20 | 1.73 | 1.15 | 9.99 | 6.36 | 8.78 | 7.80 |
| $p_m$ | *MYB23* mRNA | 9.91 | 9.99 | 4.92 | 8.19 | 1.97 | 1.41 | 9.20 | 5.71 | 8.39 | 2.19 |
| $K_c$ | WER/MYB23 complex on *CPC* transcript | 1.53 | 9.41 | 2.03 | 5.07 | 2.54 | 3.63 | 6.23 | 6.33 | 2.56 | 1.97 |
| $K_m$ | WER/MYB23 complex on *MYB23* transcript | 2.66 | 2.95 | 1.27 | 5.93 | 4.18 | 9.25 | 7.02 | 9.13 | 3.27 | 3.24 |
| $r_g$ | *GL3* by WER/MYB23 complex | 2.80 | 5.04 | 6.89 | 0.80 | 4.29 | 5.57 | 1.08 | 0.85 | 0.32 | 0.47 |
| $r_e$ | *EGL3* by WER/MYB23 complex | 2.76 | 7.08 | 1.43 | 2.94 | 8.88 | 3.33 | 6.87 | 9.57 | 8.30 | 9.74 |
| $q_G$ | GL3 protein | 3.90 | 5.83 | 2.38 | 3.53 | 8.48 | 5.29 | 0.59 | 9.26 | 7.07 | 6.76 |
| $q_E$ | EGL3 protein | 0.52 | 1.66 | 1.92 | 1.65 | 3.90 | 9.54 | 6.98 | 5.01 | 7.50 | 9.77 |
| $q_C$ | CPC protein | 4.25 | 9.22 | 6.11 | 3.77 | 5.40 | 8.54 | 5.55 | 7.24 | 5.60 | 8.92 |
| $q_W$ | WER protein | 1.63 | 2.42 | 4.91 | 3.17 | 4.29 | 3.80 | 9.42 | 4.82 | 0.98 | 4.40 |
| $q_M$ | MYB23 protein | 0.29 | 6.08 | 6.53 | 1.30 | 8.25 | 7.56 | 8.34 | 7.57 | 5.30 | 7.37 |
| $D_G$ | GL3 diffusion | 1.04 | 6.09 | 5.86 | 3.15 | 4.53 | 8.91 | 6.78 | 5.14 | 0.15 | 4.44 |
| $D_C$ | CPC diffusion | 6.71 | 5.64 | 7.77 | 9.14 | 9.21 | 6.27 | 5.68 | 7.14 | 9.13 | 1.40 |
| $n_1$ | GL3, EGL3 | 2 | 4 | 3 | 3 | 2 | 4 | 2 | 3 | 2 | 3 |
| $n_2$ | WER, MYB23 | 2 | 2 | 3 | 3 | 2 | 3 | 3 | 3 | 2 | 2 |
| $n_3$ | WER/MYB23 complex on *MYB23* | 4 | 3 | 2 | 4 | 2 | 3 | 2 | 3 | 4 | 3 |
| $n_4$ | WER/MYB23 complex on *CPC* | 3 | 3 | 3 | 2 | 2 | 2 | 2 | 3 | 3 | 3 |
| $r_{w11}$ | by CPC | 0 | 0 | 0 | 0 | 0 | 0 | 0 | 0 | 0 | 0 |
| $r_{w12}$ | by CPC complex | 7.94 | 9.44 | 8.96 | 9.93 | 0.13 | 3.71 | 6.34 | 3.58 | 2.46 | 7.17 |
| $D_E$ | diffusion | 0 | 0 | 0 | 0 | 0 | 0 | 0 | 0 | 0 | 0 |
| $S$ | SCM+ | 1 | 1 | 1 | 1 | 1 | 1 | 1 | 1 | 1 | 1 |
| $S$ | SCM- | 0 | 0 | 0 | 0 | 0 | 0 | 0 | 0 | 0 | 0 |
| $\tilde{D}_C$ | CPC import via SCM | 3.01 | 7.50 | 9.90 | 3.06 | 8.07 | 1.99 | 4.81 | 4.33 | 0.04 | 6.64 |
| $R_X$ | unknown receptor | 0.25 | 1.6e-2 | 7.8e-3 | 7.8e-3 | 3.1e-2 | 5.0e-4 | 0.25 | 1.6e-2 | 3.1e-2 | 0.25 |
| $r_{w21}$ | *WER* repression via $R_X$ | 5.37 | 3.50 | 6.87 | 3.67 | 7.96 | 1.46 | 4.58 | 8.13 | 7.84 | 7.48 |
| $C_{0_{max}}$ | maximum of random initial conditions interval | 0.5 | 6.3e-2 | 6.3e-2 | 0.13 | 0.50 | 6.3e-2 | 0.50 | 0.50 | 0.50 | 0.50 |

**Table S4-2.** Mutant data collected from the literature. [a] Measurements taken using a GL2 marker, analogous to the method used for cell type determination in our model. [b] Measurements determined from physical

examination of the cell and presence/absence of a root hair. [c] Measurements from the 'lower portion' of the root only. [d] Measurements from the 'upper portion' of the root only. [e] Numbers approximated from boxplots.

| **mutant data** | **reference** | **H position** | | **N position** | |
|---|---|---|---|---|---|
| | | **trichoblasts** | **atrichoblasts** | **trichoblasts** | **atrichoblasts** |
| *wer* | Song 2011 [1] | 99.7 ± 0.6[a] | 0.3 ± 0.6[a] | 99.3 ± 0.6[a] | 0.7 ± 0.6[a] |
| | Simon 2007 [2] | 96.8 ± 1.7[b] | 3.2 ± 1.7[b] | 88.7 ± 9.4[b] | 11.3 ± 9.4[b] |
| | Lee 2002 [3] | 95.8 ± 7.2[b] | 4.2 ± 7.2[b] | 91.3 ± 9.2[b] | 8.7 ± 9.2[b] |
| | Kwak 2007 [4] | 93.8 ± 3.8[b] | 6.2 ± 3.8[bb] | 90.6 ± 4.4[b] | 9.4 ± 4.4[b] |
| | Kang 2009 [5] | 99.8 ± 0.4[b] | 0.2 ± 0.4[b] | 97.8 ± 0.6[b] | 2.2 ± 0.6[b] |
| | Bernhardt 2003 [6] | 90.6 ± 10.0[bc] | 9.4 ± 10.0[bc] | 87.2 ± 7.3[bc] | 12.8 ± 7.3[bc] |
| | Bernhardt 2003 [6] | 97.6 ± 4.5[bd] | 2.4 ± 4.5[bd] | 99.4 ± 2.0[bd] | 0.6 ± 2.0[bd] |
| | Pietra 2015 [7] | 84.2 ± 1.7[b] | 15.8 ± 1.7[b] | 55.6 ± 1.9[b] | 44.4 ± 1.9[b] |
| | Simon 2013 [8] | 92.0[be] | 8.0[be] | 89.0[be] | 11.0[be] |
| *myb23* | Kang 2009 [5] | 97.4 ± 2.0[b] | 2.6 ± 2.0[b] | 6.0 ± 0.2[b] | 94.0 ± 0.2[b] |
| | Kang 2009 [5] | 99.2 ± 0.5[b] | 0.8 ± 0.5[b] | 4.7 ± 1.0[b] | 95.3 ± 1.0[b] |
| | Kang 2009 [5] | 91.6 ± 1.2[a] | 8.4 ± 1.2[a] | 3.8 ± 1.1[a] | 96.2 ± 1.1[a] |
| | Simon 2013 [8] | 92.0[be] | 8.0[be] | 2.0[be] | 98.0[be] |
| *wer myb23* | Simon 2013 [8] | 92.0[be] | 8.0[be] | 92.0[be] | 8.0[be] |
| *cpc try* | Simon 2007 [2] | 0.0 ± 0.0[b] | 100.0 ± 0.0[b] | 0.0 ± 0.0[b] | 100.0 ± 0.0[b] |
| | Simon 2007 [2] | 1.0 ± 0.0[a] | 99.0 ± 0.0[a] | 1.0 ± 0.0[a] | 99.0 ± 0.0[a] |
| | Kang 2013 [9] | 0.0 ± 0.0[b] | 100.0 ± 0.0[b] | 0.0 ± 0.0[b] | 100.0 ± 0.0[b] |
| | Simon 2013 [8] | 3.0[be] | 97.0[be] | 2.0[be] | 98.0[be] |
| *gl3* | Bernhardt 2003 [6] | 99.3 ± 2.1[bc] | 0.7 ± 2.1[bc] | 65.3 ± 24.3[bc] | 34.7 ± 24.3[bc] |
| | Bernhardt 2003 [6] | 96.7 ± 5.0[bd] | 3.3 ± 5.0[bd] | 6.7 ± 8.7[bd] | 93.3 ± 8.7[bd] |
| | Bernhardt 2003 [6] | 99.2 ± 2.4[bc] | 0.8 ± 2.4[bc] | 43.3 ± 13.3[bc] | 56.7 ± 12.2[bc] |
| | Bernhardt 2003 [6] | 95.8 ± 3.5[bd] | 4.2 ± 3.5[bd] | 2.5 ± 3.5[bd] | 97.5 ± 3.5[bd] |
| | Simon 2013 [8] | 93.0[be] | 7.0[be] | 44.0[be] | 56.0[be] |
| *egl3* | Bernhardt 2003 [6] | 97.3 ± 4.7[bc] | 2.7 ± 4.7[bc] | 24.0 ± 17.6[bc] | 76.0 ± 17.5[bc] |
| | Bernhardt 2003 [6] | 96.7 ± 7.1[bd] | 3.3 ± 7.1[bd] | 4.2 ± 5.0[bd] | 95.8 ± 5.0[bd] |
| | Bernhardt 2003 [6] | 92.7 ± 5.6[bc] | 7.3 ± 5.6[bc] | 27.3 ± 12.0[bc] | 72.7 ± 12.0[bc] |
| | Bernhardt 2003 [6]<br>Simon 2013 [8] | 98.3 ± 4.7[bd]<br>93.0[be] | 1.7 ± 4.7[bd]<br>7.0[be] | 1.7 ± 3.1[bd]<br>2.0[be] | 98.3 ± 3.1[bd]<br>98.0[be] |
| *gl3 egl3* | Song 2011 [1]<br>Bernhardt 2003 [6]<br>Bernhardt 2003 [6]<br>Bernhardt 2003 [6]<br>Bernhardt 2003 [6]<br>Simon 2013 [8] | 100.0 ± 0.0[a]<br>98.3 ± 4.7[bc]<br>100.0 ± 0.0[bd]<br>96.7 ± 9.4[bc]<br>100 ± 0.0[bd]<br>94.0[be] | 0.0 ± 0.0[a]<br>1.7 ± 4.7[bc]<br>0.0 ± 0.0[bd]<br>3.3 ± 9.4[bc]<br>0.0 ± 0.0[bd]<br>6.0[be] | 100.0 ± 0.0[a]<br>100.0 ± 0.0[bc]<br>96.7 ± 5.0[bd]<br>93.3 ± 12.9[bc]<br>99.2 ± 2.4[bd]<br>96.0[be] | 0.0 ± 0.0[a]<br>0.0 ± 0.0[bc]<br>3.3 ± 5.5[bd]<br>6.7 ± 12.9[bc]<br>0.8 ± 2.4[bd]<br>4.0[be] |

**Table S4-3.** Summary of solutions to mutant models within the representative cross-section topology, for comparison with mutant data in Table S4-2. 'data mean' are means calculated using data in Table S4-2. 'models mean' are means calculated using the model solutions for all ten successful parameter sets. '$P_{...}$' are the means for individual parameter sets. For the *cpc* mutant, data from the *cpc try* double mutant was used for comparison to model solutions. This is due to both the lack of single mutant data, and the fact that *try* is able to functionally substitute for *cpc* [2].

| **model** | | **H position** | | **N position** | |
|---|---|---|---|---|---|
| | | **trichoblasts** | **atrichoblasts** | **trichoblasts** | **atrichoblasts** |
| *WT* | data mean | 93.7 ± 8.4 | 6.3 ± 8.4 | 1.2 ± 3.0 | 98.8 ± 3.0 |
| | models mean | 95.9 ± 4.9 | 4.1 ± 4.9 | 0.9 ± 1.9 | 99.1 ± 1.9 |
| | $P_{3423}$ | 94.4 ± 7.6 | 5.6 ± 7.6 | 0.5 ± 2.0 | 99.5 ± 2.0 |
| | $P_{3905}$ | 100.0 ± 0.0 | 0.0 ± 0.0 | 0.0 ± 0.0 | 100.0 ± 0.0 |
| | $P_{4209}$ | 100.0 ± 0.0 | 0.0 ± 0.0 | 0.0 ± 0.0 | 100.0 ± 0.0 |
| | $P_{4634}$ | 100.0 ± 0.0 | 0.0 ± 0.0 | 0.0 ± 0.0 | 100.0 ± 0.0 |
| | $P_{6365}$ | 95.0 ± 12.2 | 5.0 ± 12.2 | 0.0 ± 0.0 | 100.0 ± 0.0 |
| | $P_{10330}$ | 100.0 ± 0.0 | 0.0 ± 0.0 | 0.0 ± 0.0 | 100.0 ± 0.0 |
| | $P_{15122}$ | 98.7 ± 3.8 | 1.3 ± 3.8 | 0.0 ± 0.0 | 100.0 ± 0.0 |
| | $P_{15348}$ | 91.3 ± 7.1 | 8.7 ± 7.1 | 3.2 ± 6.1 | 96.8 ± 6.1 |
| | $P_{18537}$ | 91.3 ± 11.5 | 8.7 ± 11.5 | 3.6 ± 6.9 | 96.4 ± 6.9 |
| | $P_{18878}$ | 96.9 ± 6.9 | 3.1 ± 6.9 | 1.8 ± 3.7 | 98.2 ± 3.7 |
| *scm* | data mean | 61.8 ± 12.9 | 38.2 ± 12.9 | 22.4 ± 15.2 | 77.6 ± 15.2 |
| | models mean | 66.0 ± 15.8 | 34.0 ± 15.8 | 31.5 ± 12.0 | 68.5 ± 12.0 |
|  | $P_{3423}$ | 61.3 ± 16.2 | 38.7 ± 16.2 | 33.2 ± 11.5 | 66.8 ± 11.5 |
|  | $P_{3905}$ | 67.5 ± 12.4 | 32.5 ± 12.4 | 35.0 ± 9.0 | 65.0 ± 9.0 |
|  | $P_{4209}$ | 73.7 ± 15.7 | 26.3 ± 15.7 | 30.5 ± 12.9 | 69.5 ± 12.9 |
|  | $P_{4634}$ | 68.7 ± 19.7 | 31.3 ± 19.7 | 26.8 ± 16.5 | 73.2 ± 16.5 |
|  | $P_{6365}$ | 61.3 ± 19.4 | 38.7 ± 19.4 | 17.7 ± 11.2 | 82.3 ± 11.2 |
|  | $P_{10330}$ | 70.0 ± 9.4 | 30.0 ± 9.4 | 26.4 ± 9.7 | 73.6 ± 9.7 |
|  | $P_{15122}$ | 56.3 ± 15.4 | 43.7 ± 15.4 | 8.2 ± 7.2 | 91.8 ± 7.2 |
|  | $P_{15348}$ | 72.5 ± 18.8 | 27.5 ± 18.8 | 22.3 ± 12.0 | 77.7 ± 12.0 |
|  | $P_{18537}$ | 58.7 ± 14.1 | 41.3 ± 14.1 | 22.7 ± 17.6 | 77.3 ± 17.6 |
|  | $P_{18878}$ | 70.0 ± 16.4 | 30.0 ± 16.4 | 27.7 ± 12.7 | 72.3 ± 12.7 |
| *wer* | data mean | 94.5 ± 3.3 | 5.5 ± 3.3 | 88.8 ± 3.9 | 11.2 ± 3.9 |
|  | models mean | 91.5 ± 5.2 | 8.5 ± 5.2 | 91.0 ± 4.6 | 9.0 ± 4.6 |
|  | $P_{3423}$ | 100.0 ± 0.0 | 0.0 ± 0.0 | 100.0 ± 0.0 | 0.0 ± 0.0 |
|  | $P_{3905}$ | 100.0 ± 0.0 | 0.0 ± 0.0 | 100.0 ± 0.0 | 0.0 ± 0.0 |
|  | $P_{4209}$ | 100.0 ± 0.0 | 0.0 ± 0.0 | 100.0 ± 0.0 | 0.0 ± 0.0 |
|  | $P_{4634}$ | 100.0 ± 0.0 | 0.0 ± 0.0 | 100.0 ± 0.0 | 0.0 ± 0.0 |
|  | $P_{6365}$ | 69.4 ± 15.4 | 30.6 ± 15.4 | 68.2 ± 14.9 | 31.8 ± 14.9 |
|  | $P_{10330}$ | 100.0 ± 0.0 | 0.0 ± 0.0 | 100.0 ± 0.0 | 0.0 ± 0.0 |
|  | $P_{15122}$ | 88.8 ± 9.9 | 11.2 ± 9.9 | 86.8 ± 8.6 | 13.2 ± 8.6 |
|  | $P_{15348}$ | 100.0 ± 0.0 | 0.0 ± 0.0 | 100.0 ± 0.0 | 0.0 ± 0.0 |
|  | $P_{18537}$ | 70.0 ± 15.4 | 30.0 ± 15.4 | 70.5 ± 13.1 | 29.5 ± 13.1 |
|  | $P_{18878}$ | 86.9 ± 11.1 | 13.1 ± 11.1 | 84.5 ± 9.8 | 15.5 ± 9.8 |
| *myb23* | data mean | 95.1 ± 0.9 | 4.9 ± 0.9 | 4.1 ± 0.6 | 95.9 ± 0.6 |
|  | models mean | 100.0 ± 0.0 | 0.0 ± 0.0 | 0.0 ± 0.0 | 100.0 ± 0.0 |
|  | all parameter sets | 100.0 ± 0.0 | 0.0 ± 0.0 | 0.0 ± 0.0 | 100.0 ± 0.0 |
| *wer myb23* | data mean | 92.0 | 8.0 | 92.0 | 8.0 |
|  | models mean | 100.0 | 0.0 | 100.0 | 0.0 |
|  | all parameter sets | 100.0 ± 0.0 | 0.0 ± 0.0 | 100.0 ± 0.0 | 0.0 ± 0.0 |
| *cpc* | data mean | 1.0 ± 0.0 | 99.0 ± 0.0 | 0.8 ± 0.0 | 99.2 ± 0.0 |
|  | models mean | 0.0 ± 0.0 | 100.0 ± 0.0 | 0.0 ± 0.0 | 100.0 ± 0.0 |
| | all parameter sets | 0.0 ± 0.0 | 100.0 ± 0.0 | 0.0 ± 0.0 | 100.0 ± 0.0 |
| *gl3* | data mean | 96.8 ± 2.6 | 3.2 ± 2.6 | 32.4 ± 10.0 | 67.6 ± 10.0 |
| | models mean | 100.0 ± 0.0 | 0.0 ± 0.0 | 0.0 ± 0.0 | 100.0 ± 0.0 |
| | all parameter sets | 100.0 ± 0.0 | 0.0 ± 0.0 | 0.0 ± 0.0 | 100.0 ± 0.0 |
| *egl3* | data mean | 95. 6 ± 4.4 | 4.4 ± 4.4 | 11.8 ± 7.5 | 88.2 ± 7.5 |
| | models mean | 97.9 ± 0.8 | 2.1 ± 0.8 | 0.6 ± 1.0 | 99.4 ± 1.0 |
| | $P_{3423}$ | 79.4 ± 8.4 | 20.6 ± 8.4 | 6.4 ± 10.3 | 93.6 ± 10.3 |
| | $P_{3905}$ | 100.0 ± 0.0 | 0.0 ± 0.0 | 0.0 ± 0.0 | 100.0 ± 0.0 |
| | $P_{4209}$ | 100.0 ± 0.0 | 0.0 ± 0.0 | 0.0 ± 0.0 | 100.0 ± 0.0 |
| | $P_{4634}$ | 100.0 ± 0.0 | 0.0 ± 0.0 | 0.0 ± 0.0 | 100.0 ± 0.0 |
| | $P_{6365}$ | 100.0 ± 0.0 | 0.0 ± 0.0 | 0.0 ± 0.0 | 100.0 ± 0.0 |
| | $P_{10330}$ | 100.0 ± 0.0 | 0.0 ± 0.0 | 0.0 ± 0.0 | 100.0 ± 0.0 |
| | $P_{15122}$ | 100.0 ± 0.0 | 0.0 ± 0.0 | 0.0 ± 0.0 | 100.0 ± 0.0 |
| | $P_{15348}$ | 100.0 ± 0.0 | 0.0 ± 0.0 | 0.0 ± 0.0 | 100.0 ± 0.0 |
| | $P_{18537}$ | 100.0 ± 0.0 | 0.0 ± 0.0 | 0.0 ± 0.0 | 100.0 ± 0.0 |
| | $P_{18878}$ | 91.9 ± 8.4 | 8.1 ± 8.4 | 8.2 ± 9.3 | 91.8 ± 9.3 |
| *gl3 egl3* | data mean | 98.2 ± 2.4 | 1.8 ± 2.4 | 97.5 ± 3.4 | 2.5 ± 3.4 |
| | models mean | 100.0 ± 0.0 | 0.0 ± 0.0 | 100.0 ± 0.0 | 0.0 ± 0.0 |
| | all parameter sets | 100.0 ± 0.0 | 0.0 ± 0.0 | 100.0 ± 0.0 | 0.0 ± 0.0 |

**Tables S4-4.** Data used to generate Figure 2C. Results for the SCM+ model solutions, solved with twenty random initial conditions for each parameter set, $P_{\dots}$, within each of the nineteen cross-section topologies found using root cross-section data, S1 File. Mean percentages and standard deviations for trichoblast cells in the H position, 'HH', atrichoblasts in the N position, 'NN', are shown in the main body of the tables. 'data' are the experimental data for *in-planta in-silico* comparison, Table 3. 'R' indicates the model results for solutions generated within the representative topology. The numbers below 'R' refer to the model results for solutions generated within topologies recorded using root cross-section data, as indexed in S1 File.

| $P_{3423}$ | HH | | NN | | $P_{3905}$ | HH | | NN | |
|---|---|---|---|---|---|---|---|---|---|
| | mean | SD | mean | SD | | mean | SD | mean | SD |
| data | 93.7 | 8.4 | 98.8 | 3.0 | data | 93.7 | 8.4 | 98.8 | 3.0 |
| R | 94.4 | 7.6 | 99.5 | 2.0 | R | 100.0 | 0.0 | 100.0 | 0.0 |
| 19 | 91.3 | 9.2 | 99.5 | 2.2 | 19 | 100.0 | 0.0 | 100.0 | 0.0 |
| 1, 16 | 98.1 | 4.6 | 100.0 | 0.0 | 1, 16 | 100.0 | 0.0 | 100.0 | 0.0 |
| 3 | 96.9 | 5.6 | 99.6 | 2.0 | 3 | 100.0 | 0.0 | 100.0 | 0.0 |
| 5 | 95.6 | 6.1 | 100.0 | 0.0 | 5 | 100.0 | 0.0 | 100.0 | 0.0 |
| 6 | 95.0 | 6.3 | 99.6 | 2.0 | 6 | 100.0 | 0.0 | 100.0 | 0.0 |
| 8 | 96.3 | 5.9 | 99.6 | 2.0 | 8 | 100.0 | 0.0 | 100.0 | 0.0 |
| 20 | 96.9 | 5.6 | 99.1 | 4.1 | 20 | 100.0 | 0.0 | 100.0 | 0.0 |
| 2 | 93.8 | 9.2 | 98.3 | 3.4 | 2 | 100.0 | 0.0 | 100.0 | 0.0 |
| 9, 14 | 96.3 | 7.1 | 97.5 | 3.9 | 9, 14 | 100.0 | 0.0 | 100.0 | 0.0 |
| 10, 11 | 95.0 | 7.5 | 98.8 | 3.1 | 10, 11 | 100.0 | 0.0 | 100.0 | 0.0 |
| 12 | 95.6 | 7.3 | 98.3 | 3.4 | 12 | 100.0 | 0.0 | 100.0 | 0.0 |
| 13 | 96.3 | 7.1 | 98.8 | 4.1 | 13 | 100.0 | 0.0 | 100.0 | 0.0 |
| 17 | 95.6 | 7.3 | 97.9 | 3.7 | 17 | 100.0 | 0.0 | 100.0 | 0.0 |
| 4 | 96.3 | 7.1 | 99.2 | 2.4 | 4 | 100.0 | 0.0 | 100.0 | 0.0 |
| 7 | 96.4 | 7.9 | 93.9 | 7.4 | 7 | 100.0 | 0.0 | 85.7 | 0.0 |
| 21 | 97.5 | 5.1 | 97.3 | 3.8 | 21 | 100.0 | 0.0 | 100.0 | 0.0 |
| 22 | 97.5 | 5.1 | 98.9 | 3.8 | 22 | 100.0 | 0.0 | 100.0 | 0.0 |
| 15 | 98.1 | 4.6 | 97.9 | 3.4 | 15 | 100.0 | 0.0 | 100.0 | 0.0 |
| 18 | 91.7 | 10.1 | 96.2 | 5.9 | 18 | 100.0 | 0.0 | 92.3 | 0.0 |

| $P_{4209}$ | HH | | NN | | $P_{4634}$ | HH | | NN | |
|---|---|---|---|---|---|---|---|---|---|
| | mean | SD | mean | SD | | mean | SD | mean | SD |
| data | 93.7 | 8.4 | 98.8 | 3.0 | data | 93.7 | 8.4 | 98.8 | 3.0 |
| R | 100.0 | 0.0 | 100.0 | 0.0 | R | 100.0 | 0.0 | 100.0 | 0.0 |
| 19 | 100.0 | 0.0 | 100.0 | 0.0 | 19 | 100.0 | 0.0 | 100.0 | 0.0 |
| 1, 16 | 100.0 | 0.0 | 100.0 | 0.0 | 1, 16 | 100.0 | 0.0 | 100.0 | 0.0 |
| 3 | 100.0 | 0.0 | 100.0 | 0.0 | 3 | 100.0 | 0.0 | 100.0 | 0.0 |
| 5 | 100.0 | 0.0 | 100.0 | 0.0 | 5 | 100.0 | 0.0 | 100.0 | 0.0 |
| 6 | 100.0 | 0.0 | 100.0 | 0.0 | 6 | 100.0 | 0.0 | 100.0 | 0.0 |
| 8 | 100.0 | 0.0 | 100.0 | 0.0 | 8 | 100.0 | 0.0 | 100.0 | 0.0 |
| 20 | 100.0 | 0.0 | 100.0 | 0.0 | 20 | 100.0 | 0.0 | 100.0 | 0.0 |
| 2 | 100.0 | 0.0 | 100.0 | 0.0 | 2 | 100.0 | 0.0 | 100.0 | 0.0 |
| 9, 14 | 100.0 | 0.0 | 100.0 | 0.0 | 9, 14 | 100.0 | 0.0 | 100.0 | 0.0 |
| 10, 11 | 100.0 | 0.0 | 100.0 | 0.0 | 10, 11 | 100.0 | 0.0 | 100.0 | 0.0 |
| 12 | 100.0 | 0.0 | 100.0 | 0.0 | 12 | 100.0 | 0.0 | 100.0 | 0.0 |
| 13 | 100.0 | 0.0 | 100.0 | 0.0 | 13 | 100.0 | 0.0 | 100.0 | 0.0 |
| 17 | 100.0 | 0.0 | 100.0 | 0.0 | 17 | 100.0 | 0.0 | 100.0 | 0.0 |
| 4 | 100.0 | 0.0 | 100.0 | 0.0 | 4 | 100.0 | 0.0 | 100.0 | 0.0 |
| 7 | 100.0 | 0.0 | 92.9 | 0.0 | 7 | 100.0 | 0.0 | 95.4 | 5.8 |
| 21 | 100.0 | 0.0 | 100.0 | 0.0 | 21 | 100.0 | 0.0 | 100.0 | 0.0 |
| 22 | 100.0 | 0.0 | 100.0 | 0.0 | 22 | 100.0 | 0.0 | 100.0 | 0.0 |
| 15 | 100.0 | 0.0 | 100.0 | 0.0 | 15 | 100.0 | 0.0 | 100.0 | 0.0 |

| 18 | 100.0 | 0.0 | 92.3 | 0.0 | 18 | 100.0 | 0.0 | 98.5 | 3.2 |
|---|---|---|---|---|---|---|---|---|---|

| $P_{6365}$ | HH | | NN | | $P_{10330}$ | HH | | NN | |
|---|---|---|---|---|---|---|---|---|---|
| | mean | SD | mean | SD | | mean | SD | mean | SD |
| data | 93.7 | 8.4 | 98.8 | 3.0 | data | 93.7 | 8.4 | 98.8 | 3.0 |
| R | 95.0 | 12.2 | 100.0 | 0.0 | R | 100.0 | 0.0 | 100.0 | 0.0 |
| 19 | 98.1 | 6.12 | 100.0 | 0.0 | 19 | 100.0 | 0.0 | 100.0 | 0.0 |
| 1, 16 | 100.0 | 0.0 | 99.6 | 2.0 | 1, 16 | 100.0 | 0.0 | 100.0 | 0.0 |
| 3 | 100.0 | 0.0 | 100.0 | 0.0 | 3 | 100.0 | 0.0 | 100.0 | 0.0 |
| 5 | 96.3 | 9.1 | 99.1 | 4.1 | 5 | 100.0 | 0.0 | 100.0 | 0.0 |
| 6 | 98.8 | 3.9 | 100.0 | 0.0 | 6 | 100.0 | 0.0 | 100.0 | 0.0 |
| 8 | 95.0 | 10.3 | 99.1 | 4.1 | 8 | 100.0 | 0.0 | 100.0 | 0.0 |
| 20 | 100.0 | 0.0 | 100.0 | 0.0 | 20 | 100.0 | 0.0 | 100.0 | 0.0 |
| 2 | 100.0 | 0.0 | 99.6 | 1.9 | 2 | 100.0 | 0.0 | 100.0 | 0.0 |
| 9, 14 | 98.8 | 5.6 | 100.0 | 0.0 | 9, 14 | 100.0 | 0.0 | 100.0 | 0.0 |
| 10, 11 | 100.0 | 0.0 | 100.0 | 0.0 | 10, 11 | 100.0 | 0.0 | 100.0 | 0.0 |
| 12 | 100.0 | 0.0 | 100.0 | 0.0 | 12 | 100.0 | 0.0 | 100.0 | 0.0 |
| 13 | 100.0 | 0.0 | 100.0 | 0.0 | 13 | 100.0 | 0.0 | 100.0 | 0.0 |
| 17 | 100.0 | 0.0 | 100.0 | 0.0 | 17 | 100.0 | 0.0 | 100.0 | 0.0 |
| 4 | 100.0 | 0.0 | 99.6 | 1.7 | 4 | 100.0 | 0.0 | 100.0 | 0.0 |
| 7 | 100.0 | 0.0 | 99.3 | 2.2 | 7 | 100.0 | 0.0 | 85.7 | 0.0 |
| 21 | 100.0 | 0.0 | 100.0 | 0.0 | 21 | 100.0 | 0.0 | 100.0 | 0.0 |
| 22 | 100.0 | 0.0 | 100.0 | 0.0 | 22 | 100.0 | 0.0 | 100.0 | 0.0 |
| 15 | 100.0 | 0.0 | 100.0 | 0.0 | 15 | 100.0 | 0.0 | 100.0 | 0.0 |
| 18 | 97.2 | 7.1 | 99.2 | 3.4 | 18 | 100.0 | 0.0 | 92.3 | 0.0 |

| $P_{15122}$ | HH | | NN | | $P_{15348}$ | HH | | NN | |
|---|---|---|---|---|---|---|---|---|---|
| | mean | SD | mean | SD | | mean | SD | mean | SD |
| data | 93.7 | 8.4 | 98.8 | 3.0 | data | 93.7 | 8.4 | 98.8 | 3.0 |
| R | 98.7 | 3.8 | 100.0 | 0.0 | R | 91.3 | 7.1 | 96.8 | 6.1 |
| 19 | 99.4 | 2.8 | 100.0 | 0.0 | 19 | 95.0 | 8.5 | 98.0 | 4.1 |
| 1, 16 | 100.0 | 0.0 | 100.0 | 0.0 | 1, 16 | 96.3 | 7.1 | 97.3 | 6.0 |
| 3 | 100.0 | 0.0 | 100.0 | 0.0 | 3 | 97.5 | 5.1 | 96.8 | 5.3 |
| 5 | 99.4 | 2.8 | 100.0 | 0.0 | 5 | 91.9 | 10.1 | 98.2 | 3.7 |
| 6 | 100.0 | 0.0 | 100.0 | 0.0 | 6 | 97.5 | 6.5 | 98.2 | 3.7 |
| 8 | 100.0 | 0.0 | 100.0 | 0.0 | 8 | 91.9 | 8.4 | 97.3 | 4.3 |
| 20 | 100.0 | 0.0 | 100.0 | 0.0 | 20 | 97.5 | 5.1 | 96.8 | 5.3 |
| 2 | 100.0 | 0.0 | 100.0 | 0.0 | 2 | 98.8 | 5.6 | 97.5 | 3.9 |
| 9, 14 | 100.0 | 0.0 | 100.0 | 0.0 | 9, 14 | 96.9 | 5.6 | 94.2 | 7.2 |
| 10, 11 | 99.4 | 2.8 | 100.0 | 0.0 | 10, 11 | 97.5 | 6.5 | 97.5 | 3.9 |
| 12 | 99.4 | 2.8 | 100.0 | 0.0 | 12 | 96.9 | 5.6 | 97.5 | 3.9 |
| 13 | 100.0 | 0.0 | 100.0 | 0.0 | 13 | 94.4 | 6.4 | 97.9 | 4.6 |
| 17 | 99.4 | 2.8 | 100.0 | 0.0 | 17 | 96.3 | 5.9 | 94.1 | 6.7 |
| 4 | 100.0 | 0.0 | 100.0 | 0.0 | 4 | 97.5 | 6.5 | 96.5 | 5.3 |
| 7 | 99.3 | 3.2 | 98.6 | 2.9 | 7 | 98.6 | 4.4 | 91.1 | 8.9 |
| 21 | 99.4 | 2.8 | 100.0 | 0.0 | 21 | 96.9 | 6.9 | 97.3 | 4.5 |
| 22 | 100.0 | 0.0 | 100.0 | 0.0 | 22 | 98.8 | 3.9 | 97.3 | 4.5 |
| 15 | 99.4 | 2.8 | 100.0 | 0.0 | 15 | 99.4 | 2.8 | 98.2 | 4.6 |
| 18 | 98.3 | 4.1 | 98.9 | 2.8 | 18 | 93.9 | 7.6 | 93.9 | 7.7 |

| $P_{18537}$ | HH | | NN | | $P_{18878}$ | HH | | NN | |
|---|---|---|---|---|---|---|---|---|---|
| | mean | SD | mean | SD | | mean | SD | mean | SD |
| data | 93.7 | 8.4 | 98.8 | 3.0 | data | 93.7 | 8.4 | 98.8 | 3.0 |
| R | 91.3 | 11.5 | 96.4 | 6.9 | R | 96.9 | 6.9 | 98.2 | 3.7 |
| 19 | 90.6 | 9.8 | 96.0 | 7.5 | 19 | 97.5 | 6.5 | 97.5 | 6.4 |
| 1, 16 | 93.1 | 8.6 | 96.4 | 6.2 | 1, 16 | 98.1 | 4.6 | 96.6 | 3.3 |
| 3 | 93.1 | 10.3 | 96.8 | 5.3 | 3 | 98.1 | 4.6 | 98.2 | 3.7 |
| 5 | 93.1 | 9.5 | 96.4 | 6.9 | 5 | 94.4 | 7.6 | 99.6 | 2.0 |
| 6 | 93.1 | 10.3 | 97.3 | 5.2 | 6 | 96.3 | 5.9 | 99.6 | 2.0 |
| 8 | 93.8 | 7.6 | 95.9 | 7.5 | 8 | 98.8 | 3.9 | 99.1 | 2.8 |
| 20 | 94.4 | 8.6 | 95.9 | 7.5 | 20 | 99.4 | 2.8 | 98.6 | 4.5 |
| 2 | 96.3 | 7.1 | 95.4 | 4.3 | 2 | 98.1 | 6.1 | 98.8 | 4.1 |
| 9, 14 | 91.3 | 8.2 | 94.2 | 6.7 | 9, 14 | 97.5 | 5.1 | 97.1 | 5.6 |
| 10, 11 | 94.4 | 7.6 | 95.4 | 6.3 | 10, 11 | 99.4 | 2.8 | 98.8 | 3.1 |
| 12 | 93.1 | 9.5 | 93.8 | 7.6 | 12 | 99.4 | 2.8 | 99.6 | 1.9 |
| 13 | 90.6 | 9.0 | 95.8 | 5.7 | 13 | 99.4 | 2.8 | 98.3 | 4.4 |
| 17 | 91.3 | 9.2 | 93.8 | 7.6 | 17 | 98.8 | 3.9 | 97.1 | 5.6 |
| 4 | 94.4 | 9.5 | 94.6 | 8.7 | 4 | 96.9 | 6.9 | 97.7 | 5.1 |
| 7 | 92.9 | 8.7 | 91.1 | 10.6 | 7 | 97.1 | 5.9 | 85.7 | 6.1 |
| 21 | 95.0 | 8.5 | 93.1 | 8.6 | 21 | 96.9 | 5.6 | 96.9 | 5.8 |
| 22 | 96.9 | 6.9 | 95.8 | 6.4 | 22 | 97.5 | 6.5 | 98.1 | 4.2 |
| 15 | 96.3 | 5.9 | 98.2 | 5.1 | 15 | 98.8 | 3.9 | 97.9 | 4.7 |
| 18 | 94.4 | 7.7 | 96.9 | 5.2 | 18 | 97.8 | 5.8 | 90.4 | 5.5 |

**Figures S3-1.** Histograms of the parameter values of the 10 successful parameter sets. Distributions for the unsuccessful parameter sets were uniform for all parameters.

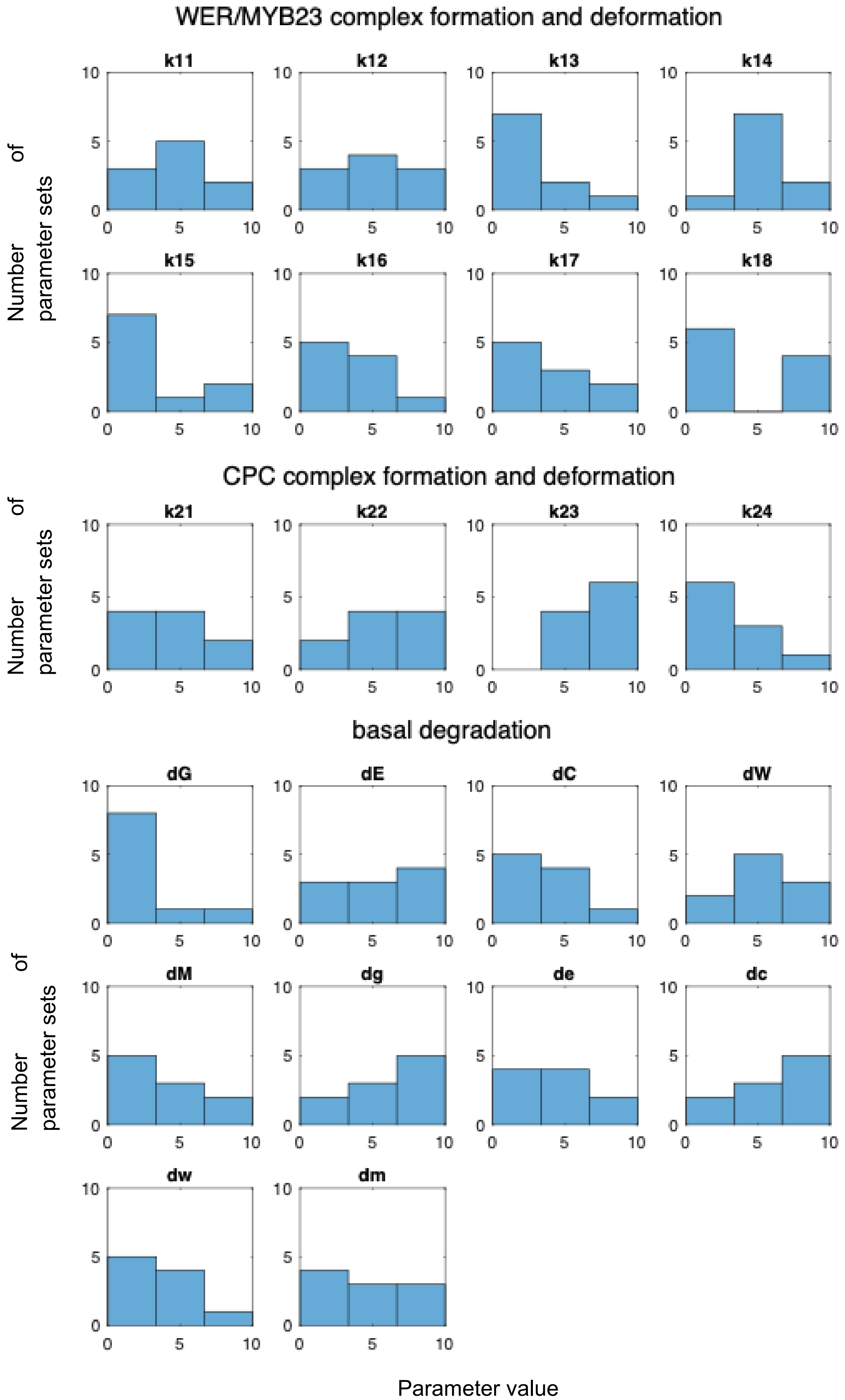

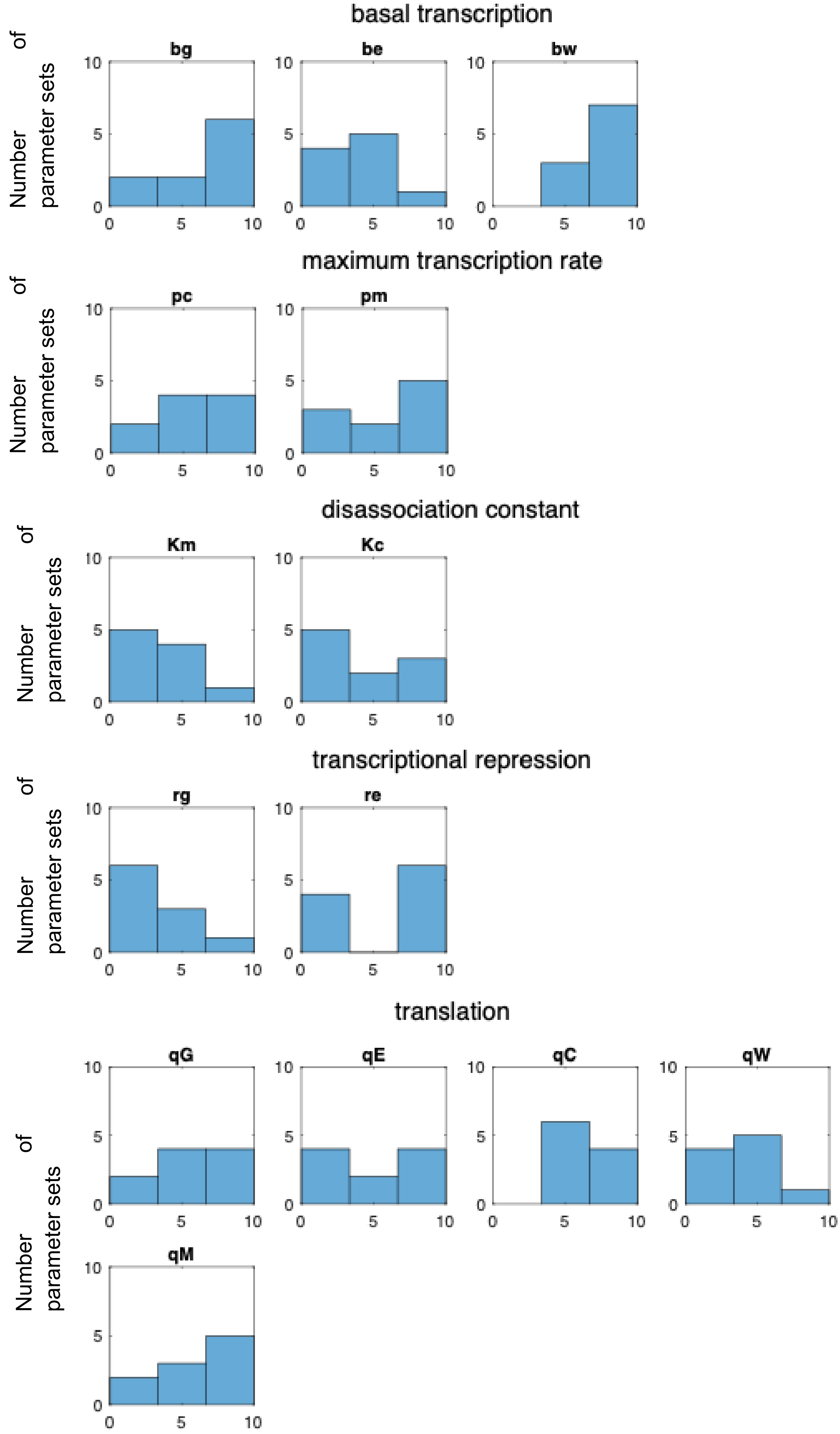

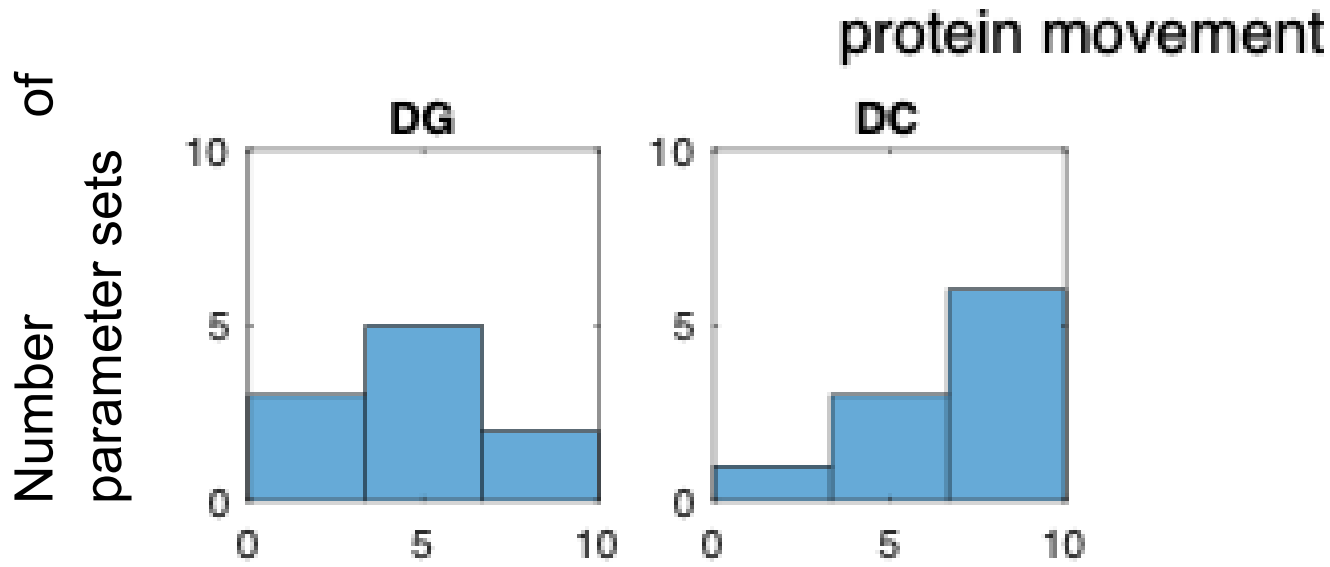

Parameters changed during mechanistic investigations
(parameters rw11, DE, and S are set to 0 in these models)

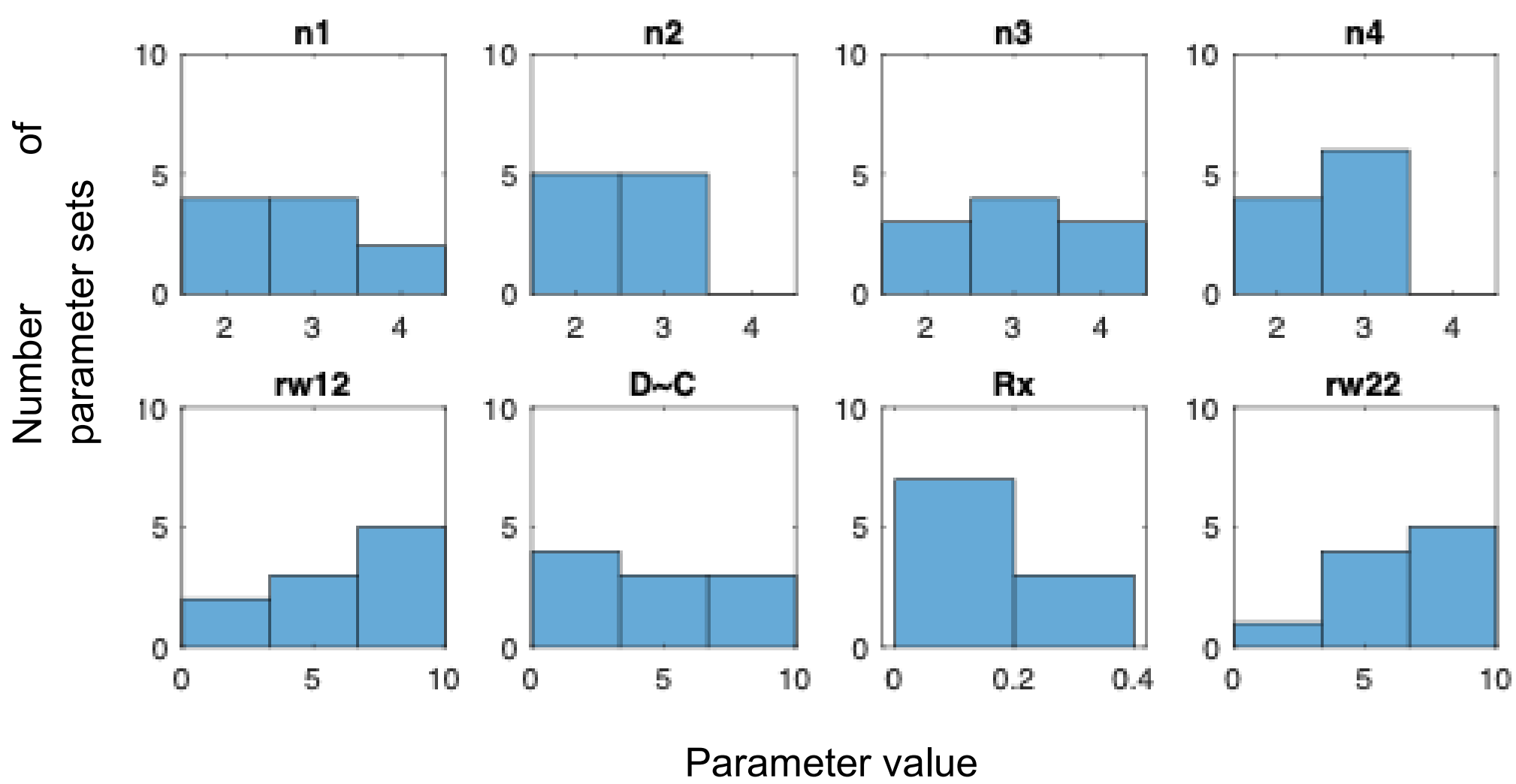

**Table S4-5.** Breakdown of the reasons for failure of the unsuccessful parameter sets, combined for the three WER regulation models.

| | **fail description** | **number of parameter sets** | **% of parameter sets** |
|---|---|---|---|
| **fails in wildtype** | homogeneous | 45,141 | 75.25 |
| | partly well defined | 2,705 | 4.51 |
| | pattern except WER | 45 | 7.5e-2 |
| | pattern except WER and wer | 3 | 5.0e-3 |
| | pattern except CPC | 514 | 0.86 |
| | pattern except GL3c | 3,240 | 5.40 |
| | pattern except MYB23 | 33 | 5.5e-2 |
| | pattern except EGL3c | 792 | 1.32 |
| | pattern except diffusing proteins | 385 | 0.64 |
| | no pattern | 3,231 | 5.39 |
| **fails in *scm* mutant** | homogeneous | 903 | 1.51 |
| | pattern does not match data | 2,998 | 5.00 |
| | **total:** | 59,990 | 100 |

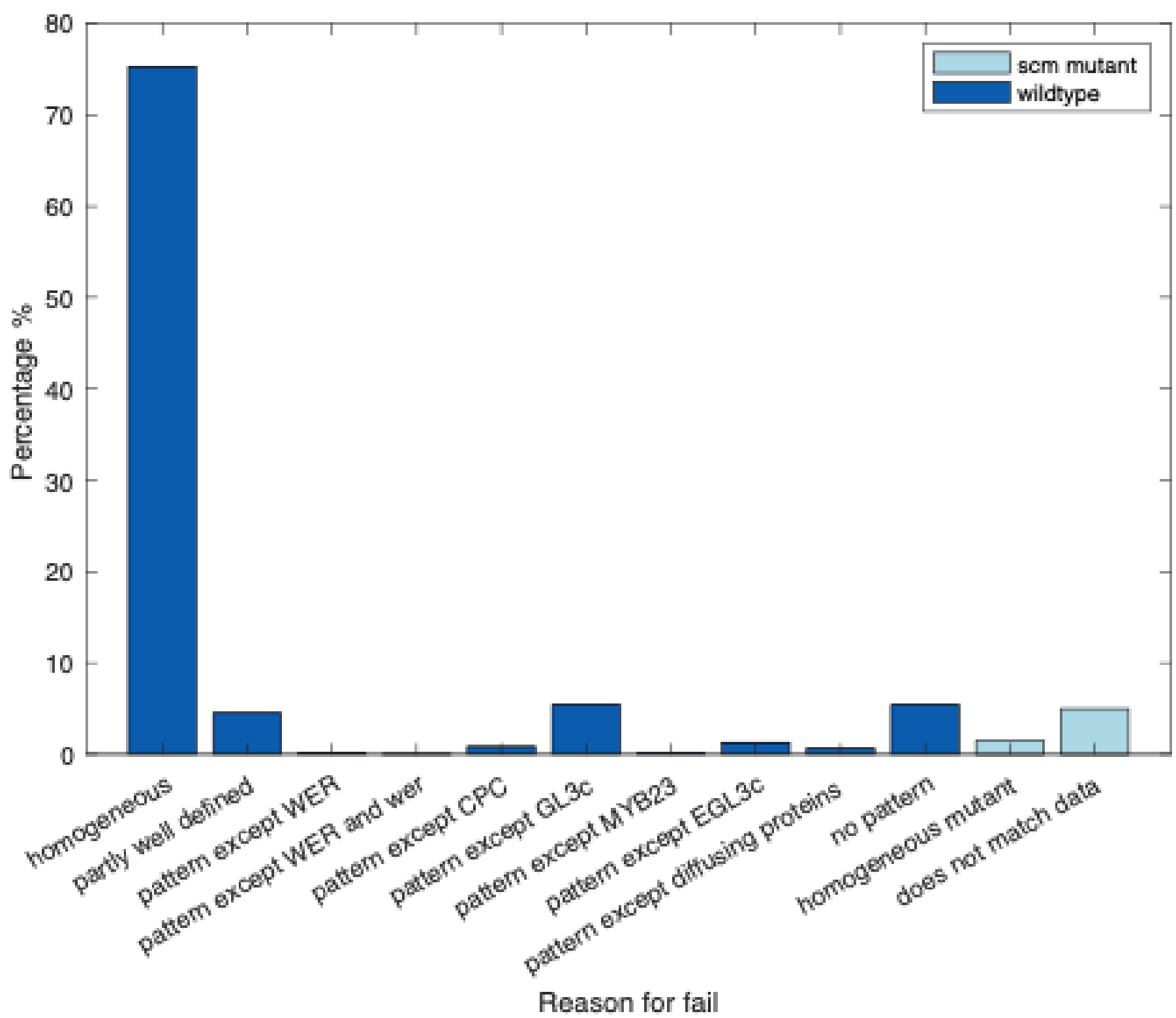

**Figure S4-2.** Breakdown of the reasons for failure of the unsuccessful parameter sets.

# File S5

Supporting information for Results Section, 'Multiple binding site reactions within the MYB23 positive feedback loop are essential for successful root hair patterning' of main manuscript.

**Table S5-1. Numbers of successful parameter sets found in the multiple binding site investigation.** Deep blue shaded cell shows the ten successful parameter sets found for the model with *WER* transcriptional repression by the cortical signal and CPC complex. The removal of multiple binding sites on the *MYB23* promotor reduced the number of successful parameter sets found (red shaded cell compared to deep blue cell). Multiple binding sites on the *MYB23* promotor only increased the number of successful parameter sets found (green shaded cells compared to blue cells). All successful parameter set numbers can be found in Table S5-2.

| number of multiple binding site reactions removed | 0 | 1 | | | 2 | | | 3 |
|---|---|---|---|---|---|---|---|---|
| | | multiple binding site removed | | | multiple binding site remaining | | | |
| ***WER* transcriptional repression** | all | *CPC* | *MYB23* | WER/MYB23 complex | *CPC* | *MYB23* | WER/MYB23 complex | none |
| cortical signal only | 0 | 0 | 0 | 0 | 0 | 0 | 0 | 0 |
| cortical signal and CPC | 0 | 0 | 0 | 0 | 0 | 3 | 0 | 0 |
| cortical signal and CPC complex | 10 | 11 | 4 | 12 | 0 | 14 | 8 | 0 |

**Table S5-2.** Successful parameter set numbers. Parameter values for all parameter sets are available on GitHub[1].

| number of multiple binding site reactions removed | 0 | 1 | | | 2 | | | 3 |
|---|---|---|---|---|---|---|---|---|
| | | multiple binding site removed | | | multiple binding site remaining | | | |
| ***WER* transcriptional repression** | all | *CPC* | *MYB23* | *WER/MYB23 complex* | *CPC* | *MYB23* | *WER/MYB23 complex* | none |
| cortical signal only | 0 | 0 | 0 | 0 | 0 | 0 | 0 | 0 |
| cortical signal and CPC | 0 | 0 | 0 | 0 | 0 | 8340<br>9625<br>19016 | 0 | 0 |

| cortical signal and CPC complex | 3423 | 984 | 578 | 4615 | 0 | 4615 | 578 | 0 |
|---|---|---|---|---|---|---|---|---|
| | 3905 | 1985 | 4209 | 5034 | | 5034 | 984 | |
| | 4209 | 2052 | 4513 | 6857 | | 6857 | 1344 | |
| | 4634 | 3905 | 6055 | 10532 | | 8558 | 1450 | |
| | 6365 | 4209 | | 10715 | | 9994 | 4209 | |
| | 10330 | 4634 | | 11615 | | 10532 | 10388 | |
| | 15122 | 4789 | | 12202 | | 10715 | 17988 | |
| | 15348 | 6365 | | 12518 | | 11615 | 19913 | |
| | 18537 | 8725 | | 15280 | | 12202 | | |
| | 18878 | 12717 | | 19016 | | 12518 | | |
| | | 18878 | | 19527 | | 14539 | | |
| | | | | 19547 | | 15280 | | |
| | | | | | | 19016 | | |
| | | | | | | 19527 | | |

# File S6

Supporting information for Results Section, 'Restricted EGL3 movement and CPC-EGL3, WER-GL3 preferential binding enable epidermal patterning' of main manuscript.

**Table S6-1. GL3 and EGL3 diffusion coefficients.** GL3 diffusion coefficient, $D_G$, EGL3 diffusion coefficient, $D_E$, and the maximum EGL3 diffusion coefficient tested which enables successful patterning, $D_{E_{MAX}}$, for each parameter set, $i$, in the investigation. The left section of the table shows the results of the parameter space analysis with $D_E > 0$, the remaining columns present the results of the investigation determining $D_{E_{MAX}}$.

| $D_E > 0$ | | | $D_{E_{MAX}}$ investigation | | | | | | | | | | | |
|---|---|---|---|---|---|---|---|---|---|---|---|---|---|---|
| $i$ | 4209 | 16189 | $i$ | 3423 | 3905 | 4209 | 4634 | 6365 | 10330 | 15122 | 15348 | 16189 | 18537 | 18878 |
| $D_G$ | 5.86 | 7.53 | $D_G$ | 1.04 | 6.09 | 5.86 | 3.15 | 4.53 | 8.91 | 6.78 | 5.14 | 7.53 | 0.15 | 4.44 |
| $D_E$ | 0.52 | 3.68 | $D_{E_{MAX}}$ | 0.92 | 0.48 | 0.66 | 0.63 | 0 | 1.41 | 0.06 | 0.13 | 3.77 | 0.14 | 0.99 |
| | | | $\alpha$ | 0.05 | 1.1 | 0.95 | 0.7 | $> 3$ | 0.8 | 2.05 | 1.6 | 0.3 | 0.05 | 0.65 |

**Table S6-2. Binding affinity ratios and the dominating complex in trichoblast and atrichoblast cells.** Data in the first three columns shows WER/GL3, MYB23/GL3 and CPC/GL3 binding affinity divided with WER/EGL3, MYB23/EGL3 and CPC/EGL3 binding affinity. If the ratio is less than 1 then the protein, WER, MYB23 or CPC, binds GL3 more strongly than EGL3, if the ratio is greater than 1 then the protein binds EGL3 more strongly than GL3. The complex which dominates in trichoblast and atrichoblast cells is also shown. Cells containing data that satisfies the relationships hypothesised as being optimal for epidermal patterning have been shaded in green.

| | WER complex | MYB23 complex | CPC complex | dominating complex | |
|---|---|---|---|---|---|
| $i$ | $\frac{k_{12}k_{13}}{k_{11}k_{14}}$ | $\frac{k_{16}k_{17}}{k_{15}k_{18}}$ | $\frac{k_{22}k_{23}}{k_{21}k_{24}}$ | trichoblast | atrichoblast |
| **3423** | 0.40 | 0.02 | 37.78 | CPC/EGL3 | MYB23/GL3 |
| **3905** | 0.50 | 1.10 | 61.37 | CPC/EGL3 | MYB23/GL3 |
| **4209** | 0.04 | 2.04 | 1.07 | CPC/GL3 | MYB23/GL3 |
| **4634** | 5.48 | 1.47 | 0.44 | CPC/GL3 | MYB23/GL3 |
| **6365** | 4.72 | 11.01 | 0.49 | CPC/GL3 | MYB23/GL3 |
| **10330** | 0.06 | 1.95 | 1.72 | CPC/EGL3 | MYB23/GL3 |
| **15122** | 1.46 | 2.64 | 9.47 | CPC/EGL3 | MYB23/EGL3 |
| **15348** | 0.41 | 382.31 | 2.81 | CPC/EGL3 | MYB23/GL3 |
| **16189** | 0.94 | 0.09 | 17.37 | CPC/EGL3 | MYB23/GL3 |
| **18537** | 1.08 | 0.45 | 4.39 | CPC/EGL3 | MYB23/GL3 |
| **18878** | 0.33 | 0.23 | 119.62 | CPC/EGL3 | MYB23/GL3 |